\documentclass[twocolumn,secnumarabic,amssymb, nobibnotes, nofootinbib, aps, prd]{revtex4-2}

\usepackage{bm}
\usepackage{amsmath,amssymb}
\usepackage{mathtools}
\usepackage[dvipdfmx]{graphicx}
\usepackage[dvipdfmx]{hyperref}
\usepackage{color}

\begin{document}


\title{Enhanced classical radiation damping of electronic cyclotron motion in the vicinity of the Van Hove singularity in a waveguide}%

\author{Yuki Goto}%
\email{goto.yuki@nifs.ac.jp}
\affiliation{National Institute for Fusion Science, National Institutes of Natural Sciences, Toki 509-5292, Japan}
\author{Savannah Garmon}%
\affiliation{Department of Physics, Osaka Metropolitan University, Sakai 599-8531, Japan}
\affiliation{Institute of Industrial Science, The University of Tokyo, Kashiwa, 277-8574, Japan}

\author{Tomio Petrosky}%
\affiliation{Center for Complex Quantum Systems, The University of Texas at Austin, Austin, 78705, USA}
\affiliation{Institute of Industrial Science, The University of Tokyo, Kashiwa, 277-8574, Japan}

\begin{abstract}
We study the damping process of electron cyclotron motion and the resulting emission in a waveguide using the classical Friedrichs model without relying on perturbation analysis such as Fermi's golden rule. A Van Hove singularity appears at the lower bound (or cut-off frequency) of the dispersion associated with each of the electromagnetic field modes in the waveguide. In the vicinity of the Van Hove singularity, we found that not only is the decay process associated with the resonance pole enhanced (amplification factor $\sim 10^4$) but the branch-point effect is also comparably enhanced. As a result, the timescale on which most of the decay occurs is dramatically shortened. Further, this suggests that the non-Markovian branch point effect should be experimentally observable in the vicinity of the Van Hove singularity. Our treatment yields a physically-acceptable solution without the problematic runaway solution that is well known to appear in the traditional treatment of classical radiation damping based on the Abraham-Lorentz equation.
\end{abstract}

\maketitle

\section{Introduction}
Cyclotron motion appears in many contexts in nature, such as auroras in Earth's magnetosphere, neutron stars in space, free electrons within metals, and magnetic confinement in fusion plasmas. Cyclotron motion plays a crucial role in astrophysics, condensed matter physics, and fusion science. As is well known, cyclotron motion is a typical example of classical radiation damping. The phenomenological Abraham-Lorentz equation has been obtained to describe radiation damping in classical mechanics \cite{Jackson98}. \\
\indent
However, the usefulness of the Abraham-Lorentz equation is limited because it leads to an unphysical solution, i.e., the so-called {\it runaway solution}, due to the force that is proportional to the third-order time derivative for the position of the charged particle. Dirac proposed a class of initial conditions that eliminate the runaway solution, however his proposal introduces non-causality \cite{Jackson98}. The inconsistency in the Abraham-Lorentz equation with the basic laws of physics arises from the fact that the field emitted in this process is calculated based on the Li\'enard–Wiechert potentials which ignore the effects of the back-reaction of the self-generated field resulting from the motion of charged particles. \\
\indent
We have two main objectives in this paper. First, we demonstrate that by using the normal mode description for the classical radiation field, as introduced in our previous paper \cite{POP03}, we can describe classical radiation damping in a manner that eliminates the inconsistency mentioned above. This is achieved through the application of a formalism parallel to that introduced by Friedrichs, which is commonly used to describe radiation damping dynamics in quantum mechanics \cite{Friedrichs48, KPPP00, PPT91, GP11}. In quantum mechanics, a decaying solution exists without any runaway behavior. Secondly, we determine the decay evolution of classical cyclotron motion and the intensity of the emitted classical fields without relying on perturbation analysis within the framework of the classical Friedrichs model introduced in this paper. This enables explicit evaluation of the decay rate of the cyclotron motion and the intensity of the emitted classical fields, even in scenarios where Fermi's golden rule may not be applicable, such as when the cyclotron frequency approaches the cut-off frequency of the waveguide.\\
\indent
In this paper, we will consider the classical system where an electron inside a rectangular electromagnetic waveguide exhibits cyclotron motion with variable frequency under the influence of a (static, uniform) external magnetic field that is applied in parallel with the length of the waveguide. It is well known that the dispersion relation of light in a waveguide has cut-off frequencies depending on each mode. For each mode, the continuous frequency spectrum of the wave has a branch point singularity at the lower bound of the spectrum in the complex frequency plane. This singular point corresponds to the Van Hove singularity that appears in solid state physics in quantum mechanics \cite{Mahan90, Van-Hove53}. In our approach, we apply the method developed in quantum mechanics as a second quantization formalism to our study \cite{KPPP00}, but now we reinterpret the annihilation and creation operators of the photon as normal modes of classical light that obey the Poisson bracket instead of the commutation relation. Since the Poisson bracket obeys the same algebra as the commutator, we can obtain an exact solution of the equation of motion for classical radiation damping. In other words, our method can be said to be a {\it classicalization} of quantum mechanics. As a result, physically meaningful solutions are obtained that avoid runaway solutions and the non-causal behavior associated with the Abraham-Lorentz equation mentioned above.\\
\indent
In early studies of cavity quantum electrodynamics, Kleppner showed that the spontaneous emission rate of an atom in a waveguide would experience a dramatic spike when the transition frequency is tuned to the cut-off (continuum band edge) of one of the waveguide modes \cite{Kleppner81}. This spike occurs as a result of the Van Hove singularity at the cut-off mode. However, Kleppner's analysis could only provide a qualitative assessment since it relies on Fermi's golden rule, which estimates the transition rate in proportion to the density of states, which diverges at the Van Hove singularity. Over a decade later, studies of emission rates in photonic band gap materials revealed more precisely the enhancement factor due to the singularity \cite{KKS94, JQ94, LNNB00}, which was later independently discovered by two of the present authors working directly on the original problem of transmission rates of atoms in waveguides \cite{PTG05}. Relying on non-perturbative methods, it was shown in \cite{PTG05} that the spontaneous emission rate $\gamma$ is modified near the cut-off such that it can be expanded in powers of $\lambda^{4/3}$, in which $\lambda \ll 1$ is the dimensionless coupling constant between the dipole moment of the atom and the field mode in the waveguide. In the context of the present work, $\lambda$ is of the order $10^{-6}$. On the other hand, when the transition frequency is far detuned from the cut-off frequency such that Fermi's golden rule can be applied (we call this the {\it Fermi region}), $\gamma$ is proportional to $\lambda^2$. Hence, the enhancement in the {\it Van Hove singularity region} compared to the Fermi region is $\lambda^{4/3}/ \lambda^2 = \lambda^{-2/3} \simeq 10^4$ times larger. This spontaneous emission process can be attributed to the appearance of the resonance pole in the complex energy plane, which is associated with Markovian (irreversible) dynamics.\\
\indent
However, in addition to the resonance pole effect, one should expect that the non-Markovian branch-point effect is also enhanced near the cut-off frequency. It is well established that the branch point effect is non-negligible at extremely long time scales and very short time scales. Under typical circumstances, the branch-point effect leads to an inverse power-law decay that only appears after many lifetimes of the exponential decay have passed \cite{Khalfin58, FGR78, RHM06, TMMS10, GPSS13, CPFSMNPO19}. On the other hand, on short time scales, the pole effect by itself would lead to pure exponential decay, which is inconsistent with the short timescale behavior, including the initial conditions. This is because there is a non-negligible deviation from exponential decay on short time scales. Thus the branch-point and the pole effect must be taken together to obtain physically sensible results. With the exception of the two experiments in Refs. \cite{RHM06, CPFSMNPO19}, the initial state is usually so depleted by the time the branch-point effect appears on the long time scale that it is extremely difficult to observe in experiment. But this effect can be enhanced near the cut-off frequency (or band edge energy) \cite{GOH21, GPSS13, JMSST05, GCV06, DBP08, GNOS19}.\\
\indent
In the present work, we extend our previous analysis to the case of cyclotronic motion inside a waveguide and further we include the branch-point effect that was previously neglected. Our results suggest that the branch-point effect indeed should be experimentally accessible in the vicinity of the Van Hove singularity at the cut-off frequency.\\
\indent
This paper is composed of six sections. In section 2, the Hamiltonian based on our physical model is provided, and the time evolution of normal modes is discussed. In section 3, the decay rate of the electron is derived. In section 4, the expressions for the field evolution associated with the pole effect and branch point effect are derived. In Section 5, their numerical calculation are discussed. The conclusion of the paper is provided in section 6.

\section{Model and Formalism}
\begin{figure}[t!]
\centering
\includegraphics[width=7.6cm]{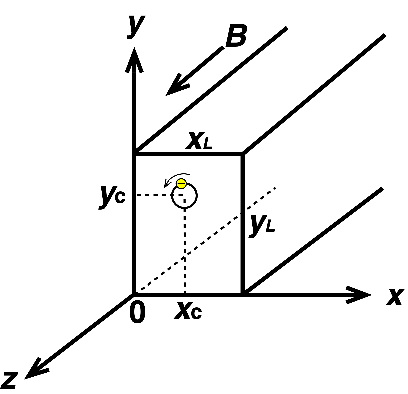}
\caption{Physical model and coordinate system. The (static, uniform) external magnetic field is applied in the positive $z$ direction along the rectangular waveguide, and the electron experiences cyclotron motion in the waveguide.}
\label{fig:rectangular_waveguide}
\end{figure}
\begin{figure}[t!]
\centering
\includegraphics[width=9cm]{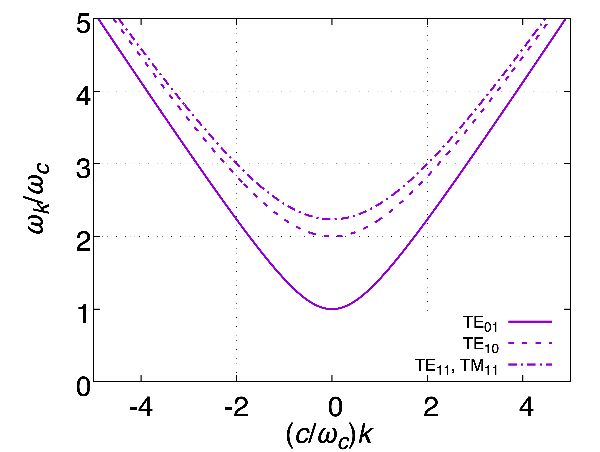}
\caption{Mode structure of the dispersion relation of light in the rectangular waveguide from the lowest mode to the fourth mode. The wave-number $k$ and frequency $\omega_{\bm k}$ are non-dimensionalized by using the cut-off frequency $\omega_c \equiv c \chi_{0,1}$ at the lowest mode of the waveguide as the unit of frequency and $c/\omega_c$ as the unit of length. As a specific example in the discussion, $(x_L,y_L)=(0.1 \,\,{\rm m}, 0.2 \,\,{\rm m})$ has been chosen, hence $\omega_c \simeq 4.7$ GHz.}
\label{fig:band_structure_rectangular}
\end{figure}

Let us consider an electron in cyclotron motion in the rectangular waveguide (see Fig.\ref{fig:rectangular_waveguide}). We assume the waveguide walls are perfect conductors. The electron couples to the electromagnetic field in the waveguide, which is infinitely long in both directions along the $z$ axis. The electron is located at the origin of the $z$ axis. The static uniform external magnetic field is applied in the positive $z$ direction. The width of the waveguide in the $x$ direction is $x_L$ and the height in the $y$ direction is $y_L$. We assume $y_L > x_L$. Also, we choose the origin for $x$ and $y$ at a corner of the rectangular cross section of the waveguide, and use a right-handed coordinate system. The cyclotron frequency can be adjusted by changing the magnitude of the magnetic field. For our numerical simulations below, we choose $(x_L,y_L)=(0.1 \,\,{\rm m}, 0.2 \,\,{\rm m})$, for which the magnitude of the magnetic field is of order $10^{-2}$ T, and the Larmor radius is of order $10^{-3}$ m. Hence the width of the waveguide is much larger than the Larmor radius, thus we can neglect the forces acting on the electron from the walls of the waveguide. These conditions induce the cyclotron emission with the emmited light in the microwave frequency regime. Under these conditions, we can write the Hamiltonian of the system for the electron as
\begin{align}
 H &= H_e + H_f,\label{hehf}
\end{align}
in which we have defined the Hamiltonian of the electron $H_e$ and Hamiltonian of the field $H_f$ as follows:
\begin{align}
 H_e &\equiv \int d \bm r \frac{1}{2m_e} \left\{\bm p + e\left[\bm A_{\rm ex}(\bm r) + \bm A_{\rm in}(\bm r)\right]\right\}^2\delta(\bm r - \bm r_e),\label{he}\\
 H_f &\equiv \int \mathrm d \bm r \left[\frac{\varepsilon_0}{2}\left|-\frac{\partial \bm A_{\rm in}(\bm r)}{\partial t}\right|^2 + \frac{1}{2\mu_0}\left|\nabla \times \bm A_{\rm in}(\bm r)\right|^2\right],\label{hf}
\end{align}
where $e$, $m_e$, $\varepsilon_0$, and $\mu_0$ are the elementary charge, the mass of the electron, the vacuum permittivity, and the vacuum permeability, respectively. Furthermore, $\bm r$ is the coordinate variable of the waveguide, and $\bm r_e$ represents the position variable of the electron. The quantity $\bm p$ is the momentum vector of the electron. We assume $\bm p$ is nonzero only in the $x$ and $y$ directions because motion in the $z$ direction does not affect the field emission. Thus, the electron rotates in the $x$-$y$ plane, and the center of the cyclotron motion is ($x_c,y_c$). Also, $\bm A_{\mathrm{ex}}(\bm r)$ is the vector potential for the (static, uniform) external magnetic field, whereas $\bm A_{\mathrm{in}}(\bm r)$ is the vector potential for the propagation modes and dressing modes around the electron in the waveguide. In the waveguide, there are two kind of propagation modes because of the boundary conditions, which are called transverse electric (TE) modes and transverse magnetic (TM) modes. For the TE (TM) modes, there is no electric (magnetic) field in the direction of propagation. Thus, the vector potential $\bm A_{\mathrm{in}}(\bm r)$ is represented by the linear combination
\begin{align}
 \bm A_{\rm in}(\bm r) &= \bm A_{\rm in}^{\rm TE}(\bm r) + \bm A_{\rm in}^{\rm TM}(\bm r). \label{vp_A}
\end{align}
By solving the sourceless Maxwell equations inside the rectangular waveguide with the boundary conditions and choosing the Coulomb gauge $\nabla \cdot \bm A_{\mathrm{in}}(\bm r) = 0$, we obtain the explicit form of the vector potentials
\begin{align}
 \bm A^{\rm TE}_{{\rm in}}(\bm r) &= \sum_{m,n}\int_{k} \frac{N_A}{\chi_{m,n}\sqrt{\omega_{\bm k}}} \left[\frac{n\pi}{y_L} F^{{\rm CS}}_{m,n}(x,y)\bm e_{x} \right. \nonumber\\
& \left. \quad\quad\quad - \frac{m\pi}{x_L} F^{{\rm SC}}_{m,n}(x,y)\bm e_{y} \right] q^{\rm TE}_{\bm k}e^{ik z} + c.c.,\label{vp_Ate}\\
 \bm A^{\rm TM}_{{\rm in}}(\bm r) &= \sum_{m,n} \int_{k} \frac{N_A c k}{\chi_{m,n}\sqrt{\omega_{\bm k}^3}} \left[\frac{m\pi}{x_L} F^{{\rm CS}}_{m,n}(x,y)\bm e_{x} \right. \nonumber\\
 &\left. + \frac{n\pi}{y_L} F^{{\rm SC}}_{m,n}(x,y)\bm e_{y} - i\frac{\chi_{m,n}^2}{k}F^{{\rm SS}}_{m,n}(x,y)\bm e_{z} \right] \nonumber\\
& \quad\quad\quad\quad\quad\quad\quad\quad\quad\quad \times q^{\rm TM}_{\bm k} e^{ik z} + c.c.,\label{vp_Atm}
\end{align}
where
\begin{align}
 N_A &= \frac{-1}{\sqrt{\pi \varepsilon_0 x_Ly_L}}, \label{NA}\\
 \chi_{m,n} &= \sqrt{\left(\frac{m\pi}{x_L}\right)^2+\left(\frac{n\pi}{y_L}\right)^2},\label{gamma_mn}
\end{align}
and
\begin{align}
 F^{{\rm CS}}_{m,n}(x,y) &= \cos\left(\frac{m\pi}{x_L}x\right) \sin\left(\frac{n\pi}{y_L}y\right),\label{wmn}\\
 F^{{\rm SC}}_{m,n}(x,y) &= \sin\left(\frac{m\pi}{x_L}x\right) \cos\left(\frac{n\pi}{y_L}y\right),\label{gmn}\\
 F^{{\rm SS}}_{m,n}(x,y) &= \sin\left(\frac{m\pi}{x_L}x\right) \sin\left(\frac{n\pi}{y_L}y\right).\label{imn}
\end{align}
The abbreviation $c.c.$ represents complex conjugate. The dispersion relation of light in the waveguide is given by
\begin{align}
 \omega_{\bm k} = c\sqrt{\chi_{m,n}^2 + k^2},\label{disp_omega}
\end{align}
where $c$, $\omega_{\bm k}$, and $k$ are the speed of light, the frequency of the field radiation, and the $z$ component of the wave number, respectively. The dispersion relation is a nonlinear function in $k$ and has cut-off frequencies $c\chi_{m,n}$ depending on each mode (see Fig.\ref{fig:band_structure_rectangular}). The quantities $q_{\bm k}^{\rm TE}$ and $q_{\bm k}^{\rm TM}$ are the normal modes of the field. The symbol $\sum_{m,n}\int_{k}$ is an abbreviation for $\sum_{m=0}^{\infty}\sum_{n=0}^{\infty}\int_{-\infty}^{\infty}dk$. The wave vector $\bm k = \left(m\pi/x_L, n\pi/y_L, k\right)$ is discrete in the $x$ and $y$ directions due to the confinement of the waveguide. The discrete pair $(m,n)$ ranges over all pairs of integers from $0$ to $\infty$, excluding $(0,0)$, while $k$ is continuous from $-\infty$ to $\infty$. We adopt a symmetric gauge as the external vector potential $\bm A_{\mathrm{ex}}(\bm r)$, which gives a static uniform magnetic field $B$ ($B>0$) in the $z$ direction:
\begin{align}
 \bm A_{\rm ex}(\bm r) = \left(-\frac{B y}{2}, \frac{B x}{2}, 0\right).\label{vp_external}
\end{align}
\indent
In this study we only consider the ${\rm TE}_{01}$ mode, which can be well separated from the other modes (see Fig.\ref{fig:band_structure_rectangular}). Then, the Hamiltonian can be approximated as follows (see Appendix A for detailed calculation):
\begin{align}
 H &\simeq \omega_1 q_1^*q_1 + \int_{-\infty}^{\infty} dk \omega_{k} q_{k}^* q_{k} \nonumber\\
 & \quad\quad\quad\quad + \lambda \int_{-\infty}^{\infty} dk \left(q_1-q_1^*\right)\left(V_k q_k-V_k^* q_k^*\right),\label{h_w_dimension}
\end{align}
where $\omega_1$ is the cyclotron frequency of the electron associated with field mode $q_1$, $\lambda$ is the dimensionless coupling constant of order $10^{-6}$ and $V_k$ is the interaction form factor, each given as follows:
\begin{align}
 \omega_1 &\equiv \frac{eB}{m_e},\label{omega1}\\
 \lambda &\equiv \sqrt{\frac{e^2\omega_c}{2m_e \varepsilon_0 c^3}},\label{lambda}\\
 V_k &\equiv i\frac{\sqrt{c^3}}{\sqrt{\pi x_Ly_L}}\sqrt{\frac{\omega_1}{\omega_c}} \frac{F^{{\rm CS}}_{0,1}(x_c,y_c)}{\sqrt{\omega_{k}}}e^{ik z_c}.\label{Vk}
\end{align}
We put $z_c=0$, assuming the electron is rotating at the origin of coordinate $z$. Since we consider only the ${\rm TE}_{01}$ mode, the bold subscript in $\omega_{\bm k}$ and $q_{\bm k}$ has been replaced with scalar $k$. This Hamiltonian has been obtained by the following two approximations, in addition to the lowest mode approximation with the ${\rm TE}_{01}$ mode. One is to neglect the field-field interaction term proportional to the square of $\bm A_{\rm in}(\bm r)$ (see Eq.\eqref{he_approx} in Appendix). The other is the dipole approximation (see Eq.\eqref{v_dipole} in Appendix). We note that virtual process terms such as $q_1 q_{k}$ and $q_1^* q_{k}^*$ are retained, i.e. the rotating wave approximation is not imposed. As a result, the field treatment is consistent with relativity, and the emission process satisfies causality. We call this Hamiltonian the classical Friedrichs model because if second quantization were performed on the normal modes of this Hamiltonian, it would become the well-known Friedrichs model in quantum mechanics \cite{KPPP00}. However, we do not perform second quantization but instead treat the normal modes as classical quantities. For the arbitrary functions $f$ and $g$ with these classical quantities as the arguments, we define the following Poisson bracket:
\begin{align}
 \{f,g\} \equiv -i\sum_{\alpha}\left(\frac{\partial f}{\partial q_{\alpha}}\frac{\partial g}{\partial q^*_{\alpha}} - \frac{\partial f}{\partial q^*_{\alpha}}\frac{\partial g}{\partial q_{\alpha}}\right),\label{PB}
\end{align}
where $\alpha$ is $1$ or $k$ and we have used a conventional discrete notation for the case of the continuous variable, i.e. $\alpha=k$ for the continuous variable, in which case the summation symbol is replaced by an integral. This definition is consistent with the definition of the Poisson brackets in terms of canonical variables (see Eq.\eqref{newpb} in Appendix A). The normal modes satisfy the Poisson brackets,
\begin{align}
 \{q_1,q^*_1\} &= -i,\label{PB_q1}\\
 \{q_k,q^*_{k'}\} &= -i\delta(k-k'),\label{PB_qk}\\
 \{q_{\alpha},q_{\beta}\} &= 0,\label{PB_q1k}
\end{align}
where $\beta$ is $1$ or $k$ with $\alpha \neq \beta$. Since the Poisson brackets satisfy the same algebra as the commutation relation, all algebraic calculations performed for our classical Friedrichs model lead to essentially the same results as the quantum case.\\
\indent
Here, we perform the non-dimensionalization of the Hamiltonian \eqref{h_w_dimension}. At first, we recognize that $eB/\chi_{m,n}^2$ has the physical dimensions of an action variable written in terms of the Hamiltonian parameters. In this study, we specifically focus on the vicinity of the Van Hove singularity associated with the lowest mode of the waveguide. Consequently, we carry out the non-dimensionalization of the Hamiltonian by using unit $J_0 \equiv eB/\chi_{0,1}^2$. Moreover, we introduce dimensionless quantities by measuring frequency in the unit of the cut-off frequency $\omega_c \equiv c \chi_{0,1}$ of the lowest waveguide mode, in which $c$ is the speed of light. Then, representing the dimensionless quantities with an overbar, we have
\begin{align}
 \bar{H} &= w_1 \bar{q}_1^* \bar{q}_1 + \int_{-\infty}^{\infty} d\kappa w_{\kappa} \bar{q}_{\kappa}^* \bar{q}_{\kappa} \nonumber\\
 & \quad\quad\quad\quad + \lambda \int_{-\infty}^{\infty} d\kappa \left(\bar{q}_1-\bar{q}_1^*\right)\left(\bar{V}_{\kappa}\bar{q}_{\kappa} - \bar{V}_{\kappa}^*\bar{q}_{\kappa}^*\right),\label{h_wo_dimen}
\end{align}
where $\bar{H} \equiv H/(\omega_c J_0)$, $\bar{q}_1 \equiv q_1/\sqrt{J_0}$, $\bar{q}_{\kappa} \equiv q_k / \sqrt{cJ_0/\omega_c}$, $\bar{V}_{\kappa} \equiv V_k/\sqrt{c\omega_c} $, $w_1 \equiv \omega_1/\omega_c$, and $w_{\kappa} \equiv \omega_k/\omega_c$. The time $t$ is also replaced by the dimensionless quantity $\tau \equiv \omega_c t$. In addition, the dispersion relation of light in the waveguide can be written as
\begin{align}
 w_{\kappa} = \sqrt{1+\kappa^2},\label{disp_wo_dimen}
\end{align}
where $\kappa \equiv (c/\omega_c) k$.\\
\indent
Since Eq.\eqref{h_wo_dimen} is a bilinear Hamiltonian, we can exactly ``diagonalize'' it as
\begin{align}
 \bar{H} = \tilde{w}_1 \bar{Q}_1^* \bar{Q}_1 + \int_{-\infty}^{\infty} d\kappa w_{\kappa} \bar{Q}_{\kappa}^* \bar{Q}_{\kappa},\label{h_diago}
\end{align}
where $\bar{Q}_{\alpha}$ are the renormalized normal modes\footnote{We note that if the cyclotron motion were to occur in free space without boundaries such as the walls of the waveguide, the renormalized normal mode of the cyclotron motion $\bar{Q}_1$ would not exist and the transformed Hamiltonian would be given by Eq.\eqref{h_diago} without the first term in the right-hand side. In this case, the Bogoliubov transformation consists of Eqs.\eqref{Qk_bog}, \eqref{q1_bog}, and \eqref{qk_bog} with $\bar{Q}_1 =0$, which was an amazing discovery by Friedrichs in \cite{Friedrichs48}. We found that the non-vanishing mode $\bar{Q}_1$ is recovered in this case due to the Van Hove singularity in the waveguide.}. The variables $(q_{\alpha},q_{\alpha}^*)$ and $(\bar{Q}_{\alpha},\bar{Q}_{\alpha}^*)$ are connected by a linear transformation,
\begin{align}
 \bar{Q}_1 &= \bar{N}_1\left[\frac{w_1 + \tilde{w}_1}{2w_1}\bar{q}_1 + \frac{w_1 - \tilde{w}_1}{2w_1}\bar{q}_1^* \right. \nonumber\\
& \left. + \lambda \int_{-\infty}^{\infty} d\kappa \left(\frac{\bar{V}_{\kappa}}{w_{\kappa}-\tilde{w}_1}\bar{q}_{\kappa} + \frac{\bar{V}_{\kappa}^*}{w_{\kappa}+\tilde{w}_1}\bar{q}_{\kappa}^*\right)\right], \label{Q1_bog}\\
\bar{Q}_{\kappa} &= \bar{q}_k - \frac{2\lambda w_1 \bar{V}_{\kappa}^*}{\bar{\xi}^{+}(w_{\kappa})} \left[\frac{w_{\kappa}+w_1}{2w_1}\bar{q}_1 - \frac{w_{\kappa}-w_1}{2w_1}\bar{q}_1^* \right. \nonumber\\
& \left. + \lambda \int_{-\infty}^{\infty} d \kappa' \left(\frac{\bar{V}_{\kappa'}}{w_{\kappa'}-w_{\kappa}-i\varepsilon}\bar{q}_{\kappa'} + \frac{\bar{V}_{\kappa'}^*}{w_{\kappa'}+w_{\kappa}}\bar{q}_{\kappa'}^*\right)\right].\label{Qk_bog}
\end{align}
where $\varepsilon$ is a positive infinitesimal. Hereafter, we leave the limit $\varepsilon \rightarrow 0$ implicit to avoid heavy notation. The function $\bar{\xi}^{\pm}$ is defined by
\begin{align}
 \bar{\xi}^{\pm}(\zeta) \equiv \zeta^2 - w_1^2 -4\lambda^2 w_1\int_{-\infty}^{\infty}d\kappa \frac{w_{\kappa}|\bar{V}_{\kappa}|^2}{\left[\zeta^2-w_{\kappa}^2\right]^{\pm}}.\label{disp_zeta}
\end{align}
Here, the $\pm$ symbol in the integrand factor $1/\left[\zeta^2-w_{\kappa}^2\right]^{\pm}$ means that the denominator is evaluated on a Riemann sheet that is analytically continued. The plus symbol indicates analytic continuation from the upper half plane to the lower half plane, and vice versa for the minus symbol. Also, $\tilde{w}_1$ is the renormalized frequency associated with the stable mode $\bar{Q}_1$, which is given by
\begin{align}
 \bar{\xi}(\tilde{w}_1) = 0.\label{fre_shifted}
\end{align}
The normalization factor $\bar{N}_1$ associated with this mode is given by
\begin{align}
 \bar{N}_1 \equiv \left[\left.\frac{1}{2w_1}\frac{d\bar{\xi}(\zeta)}{d\zeta}\right|_{\zeta=\tilde{w}_1}\right]^{-\frac{1}{2}}.\label{N1}
\end{align}
We refer to the above transformation as the classical Bogoliubov transformation (see Appendix B for detailed calculation). The inverse transformation is given by
\begin{align}
 \bar{q}_1 &= \bar{N}_1\left(\frac{\tilde{w}_1 + w_1}{2w_1}\bar{Q}_1 - \frac{\tilde{w}_1 - w_1}{2w_1}\bar{Q}_1^*\right) \nonumber\\
& - \lambda \int_{-\infty}^{\infty}d \kappa \left[\frac{(w_{\kappa}+w_1)\bar{V}_{\kappa}}{\bar{\xi}^-(w_{\kappa})}\bar{Q}_{\kappa} + \frac{(w_{\kappa}-w_1)\bar{V}_{\kappa}^*}{\bar{\xi}^+(w_{\kappa})}\bar{Q}_{\kappa}^*\right],\label{q1_bog}\\
 \bar{q}_{\kappa} &= \bar{Q}_{\kappa} - \lambda \bar{N}_1 \bar{V}_{\kappa}^*\left(\frac{\bar{Q}_1}{\tilde{w}_1 - w_{\kappa}} + \frac{\bar{Q}_1^*}{\tilde{w}_1 + w_{\kappa}}\right)\nonumber\\
& + 2\lambda^2w_1 \bar{V}_{\kappa}^*\int_{-\infty}^{\infty} d \kappa' \left[\frac{\bar{V}_{\kappa'}}{\bar{\xi}^-(w_{\kappa'})(w_{\kappa'}-w_{\kappa}-i\varepsilon)}\bar{Q}_{\kappa'} \right. \nonumber\\
& \left. \quad\quad\quad\quad\quad\quad\quad\quad + \frac{\bar{V}_{\kappa'}^*}{\bar{\xi}^+(w_{\kappa'})(w_{\kappa'}+w_{\kappa})}\bar{Q}_{\kappa'}^*\right].\label{qk_bog}
\end{align}
Corresponding to Eq.\eqref{PB_q1} through Eq.\eqref{PB_q1k}, the renormalized normal modes $(\bar{Q}_{\alpha},\bar{Q}_{\alpha}^*)$ satisfy the Poisson brackets,
\begin{align}
 \{\bar{Q}_1,\bar{Q}^*_1\} &= -i,\label{PB_Q1}\\
 \{\bar{Q}_{\kappa},\bar{Q}^*_{\kappa'}\} &= -i\delta(\kappa-\kappa'),\label{PB_Qk}\\
 \{\bar{Q}_{\alpha},\bar{Q}_{\beta}\} &= 0.\label{PB_Q1k}
\end{align}
where the Poisson bracket can be expressed by replacing $q_{\alpha}$ by $\bar {Q}_{\alpha}$ in Eq.\eqref{PB}.\\
\indent
Performing the integral over $\kappa$ in Eq.\eqref{disp_zeta}, we find
\begin{align}
 \bar{\xi}^{\pm}(\zeta) = \zeta^2 - w_1^2 \pm \frac{\lambda^2G^2w_1}{\sqrt{1-\zeta^2}},\label{disp_zeta_int}
\end{align}
where we have defined the dimensionless constant
\begin{align}
 G^2 \equiv \frac{4c^2}{x_Ly_L\omega_c^2}\sin^2\left(\frac{\pi}{y_L}y_c\right),\label{G_sq}
\end{align}
for which $G \sim 1$. Furthermore, the normalization factor \eqref{N1} can be evaluated by using Eq.\eqref{disp_zeta_int}, which gives
\begin{align}
 \bar{N}_1 = \left[\frac{\tilde{w}_1}{w_1} + \frac{\lambda^2G^2\tilde{w}_1}{2\left(1-\tilde{w}_1^2\right)^{3/2}}\right]^{-\frac{1}{2}},
\end{align}
where $\bar{N}_1 \to 0$ in the limit $w_1 \to \infty$, because $\tilde{w}_1 \to 1$ in this limit (see Fig.\ref{fig:exceptional_point}(a)).\\
\indent
Since the Hamiltonian \eqref{h_wo_dimen} was diagonalized as Eq.\eqref{h_diago}, we can obtain the time evolution of the renormalized normal mode $\bar{Q}_{\alpha}$ by Hamilton's equation of motion. Hamilton's equation of motion for $\bar{Q}_{\alpha}$ is
\begin{align}
 i\frac{d\bar{Q}_{\alpha}}{d\tau} &= -L_H\bar{Q}_{\alpha},\label{hem}
\end{align}
where $L_H$ is the Liouville operator (Liouvillian), which is defined by the Poisson bracket with the Hamiltonian and is a generator of the time evolution in classical mechanics,
\begin{align}
 L_H f \equiv i\{H,f\}.\label{Lio}
\end{align}
The imaginary number $i$ in front of the right-hand side of Eq.\eqref{Lio} is introduced so the Liouvillian $L_H$ is a Hermitian operator in the Hilbert space. Substituting the Hamiltonian \eqref{h_diago} into Eq.\eqref{Lio}, one can see that $\bar{Q}_{\alpha}$ is an eigenfunction of $L_H$,
\begin{align}
 -L_H\bar{Q}_{\alpha} &= \tilde{w}_{\alpha} \bar{Q}_{\alpha},\label{Lio_eigen}
\end{align}
where $\tilde{w}_1 \neq w_1$ is the renormalized frequency of  the renormalized normal mode of the cyclotron motion, and $\tilde{w}_{\kappa} = w_{\kappa}$ is the frequency of the the renormalized field mode. By Eqs.\eqref{hem} and \eqref{Lio_eigen}, the time evolution of $\bar{Q}_{\alpha}$ becomes
\begin{align}
 \bar{Q}_{\alpha} &= \bar{Q}_{\alpha}(0)e^{-i\tilde{w}_{\alpha} \tau},\label{Q_tim_evo}
\end{align}
Assuming $\bar{q}_{\kappa}=0$ at $\tau=0$, the initial conditions $\bar{Q}_1(0)$ and $\bar{Q}_{\kappa}(0)$ are given through Eqs.\eqref{Q1_bog} and \eqref{Qk_bog} by
\begin{align}
 \bar{Q}_1(0) &= \bar{N}_1\left[\frac{w_1+\tilde{w}_1}{2w_1}\bar{q}_1(0) + \frac{w_1-\tilde{w}_1}{2w_1}\bar{q}_1^*(0)\right],\label{Q1_inti}\\
 \bar{Q}_{\kappa}(0) &= - \frac{2\lambda w_1 \bar{V}_{\kappa}^*}{\bar{\xi}^{+}(w_{\kappa})}\left[\frac{w_{\kappa}+w_1}{2w_1}\bar{q}_1(0)-\frac{w_{\kappa}-w_1}{2w_1}\bar{q}_1^*(0)\right].\label{Qk_inti}
\end{align}

\section{Decay Rate of the Cyclotron Motion}
\begin{figure}[t!]
\centering
\includegraphics[width=\linewidth]{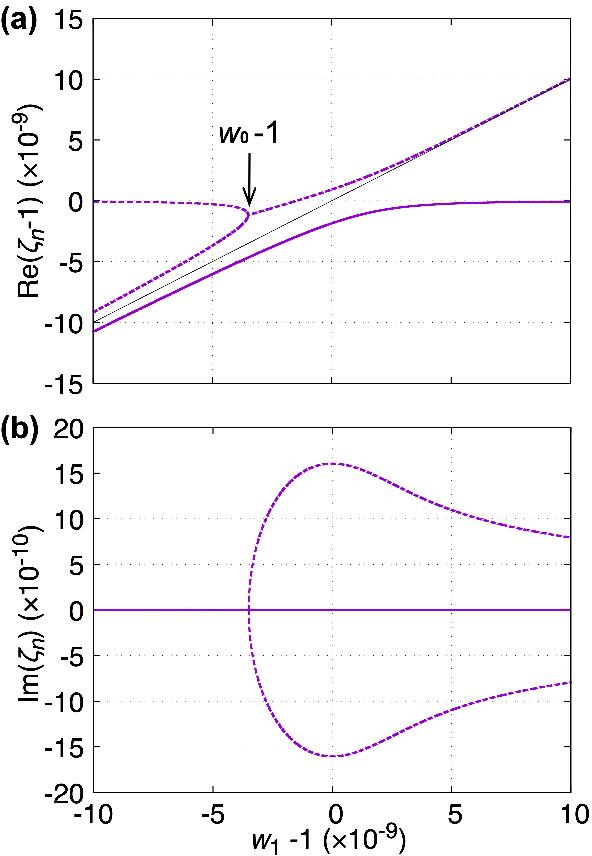}
\caption{Pole locations in the vicinity of the Van Hove singularity. (a) is the real part, (b) is the imaginary part. The solid lines are for $\zeta_0$, the dashed lines are for $\zeta_1$ and $\zeta_2$, respectively. The imaginary part of the $\zeta_1$ in (b) provides the decay rate of the cyclotron motion. The maximum decay rate on the Van Hove singularity $(w_1-1=0)$ is $10^4$ times larger compared to the Fermi region $(w_1 \gg 1)$ which is located to the far right, outside of the graph.}
\label{fig:exceptional_point}
\end{figure}
The electron mode $\bar{q}_1$ in Eq.\eqref{q1_bog} consists of two terms,
\begin{align}
 \bar{q}_1(\tau) = \bar{q}_{1,s}(\tau) + \bar{q}_{1,d}(\tau),
\end{align}
which are the discrete component associated with $\bar{Q}_1$, given by
\begin{align}
 \bar{q}_{1,s}(\tau) &\equiv \bar{N}_1\left[\frac{\tilde{w}_1 + w_1}{2w_1}\bar{Q}_{1}(0)e^{-i\tilde{w}_{1} \tau} \right.\nonumber\\
 &\quad\quad\quad\quad\quad\quad\quad\quad \left. - \frac{\tilde{w}_1 - w_1}{2w_1}\bar{Q}_{1}^*(0)e^{i\tilde{w}_{1} \tau}\right],\label{q1s}
\end{align}
and the field component
\begin{align}
  \bar{q}_{1,d}(\tau) &\equiv - \lambda \int_{-\infty}^{\infty}d \kappa \left[\frac{(w_{\kappa}+w_1)\bar{V}_{\kappa}}{\bar{\xi}^-(w_{\kappa})}\bar{Q}_{\kappa}(0)e^{-iw_{\kappa} \tau} \right.\nonumber\\
&\quad\quad\quad\quad\quad \left.+ \frac{(w_{\kappa}-w_1)\bar{V}_{\kappa}^*}{\bar{\xi}^+(w_{\kappa})}\bar{Q}_{\kappa}^*(0)e^{iw_{\kappa} \tau}\right].\label{q1d}
\end{align}
As can be seen when substituting $\bar{Q}_{1}(0)$ into Eq.\eqref{q1s}, $\bar{q}_{1,s}$ simply gives a stable oscillation without decay. Therefore, the cyclotron motion inside the waveguide never stops even as $\tau \to +\infty$. This means the electron experiences perpetual acceleration without emitting radiation even as $\tau \to +\infty$. This is possible because of the Van Hove singularity.\\
\indent
To evaluate $\bar{q}_{1,d}$, we substitute $\bar{Q}_{\kappa}(0)$ into Eq.\eqref{q1d} and change the integration variable from $\kappa$ to $w_{\kappa}$, to obtain
\begin{align}
 \bar{q}_{1,d}(\tau) &=2\lambda^2 \int_{1}^{\infty}d w_{\kappa} \frac{w_{\kappa}}{\sqrt{w_{\kappa}^2 -1}} \frac{\left|\bar{V}_{\kappa}\right|^2}{\left|\bar{\xi}^+(w_{\kappa})\right|^2} \nonumber\\
 &\times \left[(w_{\kappa}+w_1) h(w_{\kappa})e^{-iw_{\kappa} \tau} \right. \nonumber\\
 &\quad\quad\quad\quad\quad\quad \left.+ (w_{\kappa}-w_1)h^*(w_{\kappa})e^{iw_{\kappa} \tau}\right],\label{q1d_wk}
\end{align}
where
\begin{align}
 h(w_{\kappa}) &\equiv w_1{\rm{Re}}[\bar{q}_1(0)] + i w_{\kappa}{\rm{Im}}[\bar{q}_1(0)],\label{hwk}
\end{align}
and
\begin{align}
 \left|\bar{\xi}^+(w_{\kappa})\right|^2 = \bar{\xi}^+(w_{\kappa}) \bar{\xi}^-(w_{\kappa}).
\end{align}
Here, the factor $d\kappa/dw_{\kappa} = w_{\kappa}/\sqrt{w_{\kappa}^2 -1}$ is the density of states, which diverges at the lower limit of the integral. This divergence is the Van Hove singularity.\\
\indent
As can be seen, $\bar{q}_{1,d}$ gives the time decay of electron motion because the denominetor in Eq.\eqref{q1d_wk} has poles in the complex frequency plane. By putting the denominetor in Eq.\eqref{q1d_wk} to zero, we obtain a double cubic equation as follows:
\begin{align}
 \left(\zeta_n^2\right)^3 - (2w_1^2+1)\left(\zeta_n^2\right)^2 &+ (2w_1^2+w_1^4)\zeta_n^2 \nonumber\\
 & \quad - w_1^4 + \lambda^4G^4w_1^2 = 0,\label{cubic}
\end{align}
where we have introduced the dimensionless variable $\zeta_n^2$. The general solutions of Eq.\eqref{cubic} are given by the standard method to solve the cubic equation as
\begin{align}
 \zeta_n^2 = e^{\frac{2}{3}ni\pi}\alpha_{w_{1+}}^{\frac{1}{3}} + e^{-\frac{2}{3}ni\pi}\alpha_{w_{1-}}^{\frac{1}{3}} + d_{w_1},\label{zn}
\end{align}
where
\begin{align}
 \alpha_{w_{1\pm}} \equiv \frac{-q_{w_1}\pm \sqrt{q^2_{w_1}+4p^3_{w_1}}}{2},\label{alpha_pm}
\end{align}
and
\begin{align}
 p_{w_1} &\equiv -\frac{1}{3^2}(w_1^2-1)^2,\label{pw1}\\
 q_{w_1} &\equiv \frac{2}{3^3}(w_1^2-1)^3+w_1^2\lambda^4G^4,\label{qw1}\\
 d_{w_1} &\equiv \frac{1}{3}(2w_1^2+1).\label{dw1}
\end{align}
Here $n=0$, $1$ or $2$. By taking the square root of eq.\eqref{zn}, we obtain three solutions $\zeta_n$ with ${\rm Re}[\zeta_n] > 0$. Additionally, the three solutions with ${\rm Re}[\zeta_n] < 0$ are given by putting a minus sign to the solutions with ${\rm Re}[\zeta_n] > 0$. In the following, we show the behaviour of the three solutions with ${\rm Re}[\zeta_n] > 0$. The solution of this equation labeled $\zeta_0$ is real for all values of $w_1$ and corresponds to the bound state discussed previously. The solutions $\zeta_1$ and $\zeta_2$ can be real or complex-valued, depending on the value of $w_1$ as described below.\\
\indent
Fig.\ref{fig:exceptional_point} shows the behavior of the solutions $\zeta_n$ in the vicinity of the cut-off frequency (i.e., in the vicinity of the Van Hove singularity) in the waveguide. Fig.\ref{fig:exceptional_point}(a) represents the real part of the solutions, while (b) represents the imaginary part. In both figures, the solid line corresponds to the bound state $\zeta_0$, while the dashed lines represent the two complex solutions $\zeta_1$ and $\zeta_2$. The complex solution with negative imaginary part corresponds to the resonance state that breaks time symmetry.\\
\indent
Moreover, one can see in Fig.\ref{fig:exceptional_point}(a) that there exists a point $w_0-1$ where two real solutions coalesce before turning into a resonance and its partner anti-resonance state with complex eigenvalues. This point is known as an exceptional point \cite{GOH21}. The properties of the system near the exceptional point are interesting, but we will focus in this paper more narrowly on the Van Hove singularity itself.\\
\indent
Exactly at the Van Hove singularity, we can obtain $\zeta_{1}^2$ as the exact solution of the cubic equation \eqref{cubic} after putting $w_1=1$,
\begin{align}
 \zeta_{1}^2 = 1 + \frac{1}{2}\lambda^{\frac{4}{3}}G^{\frac{4}{3}} - i\frac{\sqrt{3}}{2}\lambda^{\frac{4}{3}}G^{\frac{4}{3}},\label{Puiseux_zeta_1^2}
\end{align}
where $G$ is the dimensionless constant with typical order $1$ as shown in Eq.\eqref{G_sq}. The resonance state $\zeta_{1}$ at the Van Hove singularity is then given by taking the square root of Eq.\eqref{Puiseux_zeta_1^2}, then we have
\begin{align}
 \zeta_{1} = 1 + \frac{1}{4}\lambda^{\frac{4}{3}}G^{\frac{4}{3}} - i\frac{\sqrt{3}}{4}\lambda^{\frac{4}{3}}G^{\frac{4}{3}} + \mathcal{O}\left(\lambda^{\frac{8}{3}}\right).\label{Puiseux}
\end{align}
This expression can also be obtained by Puiseux expansion in terms of $\lambda$ \cite{GOH21}. The imaginary part of this expression gives the decay rate at the Van Hove singularity. From the form of Eq.\eqref{Puiseux}, it is clear that this result cannot be obtained by the usual perturbation analysis with the series expansion in terms of $\lambda$ as in the case of Fermi's golden rule that gives the decay rate proportional to $\lambda^2$. We can see that the decay rate in the vicinity of the Van Hove singularity is magnified by a factor $\lambda^{-2/3}(=\lambda^{4/3}/\lambda^{2})$ compared with the Fermi region. For the electron cyclotron motion, this magnification factor is about $10^4$.\\
\indent
Regarding this discussion, we elaborate further on the point that $\tilde{w}_1$ and $w_{\kappa}$ represent the real eigenvalues in the first Riemann sheet and the complex eigenvalues in the second Riemann sheet of the Liouvillian, respectively. Remarkably, $\bar{Q}_1$ does not experience time decay, whereas $\bar{Q}_{\kappa}$ experiences time decay. As briefly mentioned in Footnote 1, the existence of the Van Hove singularity allows for this non-decaying mode. Consequently, there is a scenario where cyclotron motion can persist without emitting light. In other words, in classical systems, there are instances where an electron can undergo accelerated motion without emitting light and continue its motion without decay.

\section{Time Evolution of the Field Mode}
\begin{figure}[t!]
\centering
\includegraphics[width=\linewidth]{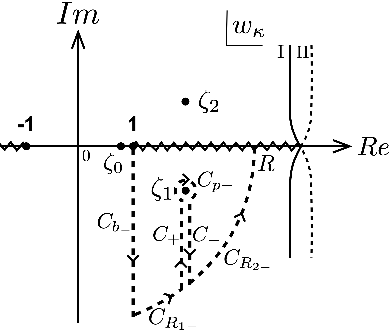}
\caption{Pole locations on the complex frequency plane. $\zeta_0$ is located in the first Riemann sheet. $\zeta_1$ and $\zeta_2$ are located in the second Riemann sheet.}
\label{fig:residue_paper}
\end{figure}
The field modes $\bar{q}_{\kappa}$ in Eq.\eqref{qk_bog} consists of three terms,
\begin{align}
 \bar{q}_{\kappa}(\tau) = \bar{q}_{\kappa,p}(\tau) + \bar{q}_{\kappa,s}(\tau) + \bar{q}_{\kappa,d}(\tau),\label{qk_three_term}
\end{align}
where
\begin{align}
 \bar{q}_{\kappa,p}(\tau) &\equiv \bar{Q}_{\kappa}(0)e^{-iw_{\kappa} \tau},\label{qkp}\\
 \bar{q}_{\kappa,s}(\tau) &\equiv - \lambda \bar{N}_1 \bar{V}_{\kappa}^*\left[\frac{\bar{Q}_{1}(0)e^{-i\tilde{w}_{1} \tau}}{\tilde{w}_1 - w_{\kappa}} + \frac{\bar{Q}_{1}^*(0)e^{i\tilde{w}_{1} \tau}}{\tilde{w}_1 + w_{\kappa}}\right],\label{qks}\\
 \bar{q}_{\kappa,d}(\tau) &\equiv 2\lambda^2w_1 \bar{V}_{\kappa}^*\int_{-\infty}^{\infty} d \kappa' \left[\frac{\bar{V}_{\kappa'}\bar{Q}_{\kappa'}(0)e^{-iw_{\kappa'} \tau}}{\bar{\xi}^-(w_{\kappa'})(w_{\kappa'}-w_{\kappa}-i\varepsilon)} \right. \nonumber\\
&\left. \quad\quad\quad\quad\quad\quad\quad\quad + \frac{\bar{V}_{\kappa'}^* \bar{Q}_{\kappa'}^*(0)e^{iw_{\kappa'} \tau}}{\bar{\xi}^+(w_{\kappa'})(w_{\kappa'}+w_{\kappa})}\right]\label{qkd}.
\end{align}
The term $\bar{q}_{\kappa,p}$ describes the free propagation of the field inside the waveguide. On the other hand, $\bar{q}_{\kappa,s}$ and $\bar{q}_{\kappa,d}$ describe the steady oscillation and the decaying components of the field around the electron, respectively.\\
\indent
The focus of our interest is a detailed analysis of the propagating wave emitted toward the far distance from the electron in cyclotron motion, which is denoted as $\bar{q}_{\kappa,p}$. Therefore, we do not further discuss the bound fields around the electron $\bar{q}_{\kappa,s}$ and $\bar{q}_{\kappa,d}$ in the present work.\\
\indent
For the ${\rm TE}_{01}$ mode which we consider here, the vector potential \eqref{vp_Ate} is approximated as
\begin{align}
 A_p(Z, \tau) &\simeq A_c \int_{-\infty}^{\infty} d\kappa \frac{1}{\sqrt{w_{\kappa}}} \bar{q}_{\kappa,p}(\tau)e^{i\kappa Z} + c.c.,\label{vp_Ap}
\end{align}
where $Z$ is the dimensionless coordinate variable on the $z$ direction defined by $Z \equiv (\omega_c/c)z$ and we have introduced the notation $A_p(Z, \tau)$ as an abbreviation for $A^{{\rm TE}_{01}}_{{\rm in},x}(Z,\tau)$. The quantity $A_c$ is the amplitude with the dimension of a vector potential:
\begin{align}
 A_c \equiv \sqrt{\frac{J_0}{c}}F^{\rm CS}_{0,1}(x_c,y_c)N_A.\label{Ac}
\end{align}
We note that $(x,y)$ in Eq.\eqref{Ac} has been replaced with the rotation center of the electron $(x_c,y_c)$, because of the dipole approximation.\\
\indent
In the following part of this paper, we calculate the time evolution at $Z=0$. This is because, once the field is emitted from the cyclotron motion, the field can propagate freely inside the waveguide. Thus, if one wishes, it is possible to calculate the propagation of the emitted field by using the linear wave equation with a given source that is the time  evolution of the field at $Z=0$.\\
\indent
By substituting Eq.\eqref{qkp} with initial condition \eqref{Qk_inti} into Eq.\eqref{vp_Ap}, the electric field at $Z=0$ is given by
\begin{align}
 E_{p}(0,\tau) &= -\omega_c \frac{\partial A_p(0,\tau)}{\partial \tau}\nonumber\\
 &=\lambda E_c  \int_{-\infty}^{\infty} d\kappa \frac{h(w_{\kappa})}{\bar{\xi}^+(w_{\kappa})} e^{-iw_{\kappa} \tau} + c.c.,\label{Ep}
\end{align}
where $h(w_{\kappa})$ is given in Eq.\eqref{hwk} and the factor $E_c$ in Eq.\eqref{Ep} is a quantity with the dimension of the electric field defined as
\begin{align}
 E_c &\equiv \frac{-2 \sqrt{w_1} c F^{\rm CS}_{0,1}(x_c,y_c)A_c}{\sqrt{\pi x_Ly_L}}\nonumber\\
 &=\frac{\omega_c^2}{2\pi}\sqrt{\frac{w_1 J_0}{c^3\varepsilon_0}}G^2.\label{Ec}
\end{align}
By changing the integration variable from $\kappa$ to $w_{\kappa}$, we have
\begin{align}
 E_{p}(0,\tau) = \lambda E_c \int_{1}^{\infty} dw_{\kappa} \tilde{g}^{+}(w_{\kappa}) e^{-iw_{\kappa} \tau} + c.c..\label{Ep_wk}
\end{align}
where
\begin{align}
 \bar{g}^{+}(w_{\kappa}) &\equiv \frac{w_{\kappa}}{\sqrt{w_{\kappa}^2 -1 } } \frac{h(w_{\kappa})}{\bar{\xi}^{+}(w_{\kappa})}\nonumber\\
 &= \frac{w_{\kappa}w_1{\rm{Re}}[\bar{q}_1(0)] + i w_{\kappa}^2 {\rm{Im}}[\bar{q}_1(0)]}{\sqrt{w_{\kappa}^2-1}\left(w_{\kappa}^2-w_1^2\right) - i \lambda^2 G^2 w_1}.\label{g_tilde}
\end{align}
Here again, $d\kappa/dw_{\kappa} = w_{\kappa}/\sqrt{w_{\kappa}^2 -1}$ is the density of states, which diverges at the lower limit of the integral.\\
\indent
We note that the function $\bar{g}^{+}$ does not diverge at the lower limit of the integration. This is because $\bar{\xi}^{+}$ given in the denominator of Eq.\eqref{disp_zeta_int} also has a divergence at the lower limit of integration, thus the product of the density of states $d\kappa/dw_{\kappa}$ and the function in the denominator $\bar{\xi}^{+}$ suppresses the divergence of the integrand. However, the lower limit of the integral is still a singular point. This is because the denominator of $\bar{g}^{+}$ includes the branch point due to the density of states.\\
\indent
Putting the denominator of $\bar{g}^{+}$ to zero in Eq.\eqref{Ep_wk}, we obtain the same cubic equation in Eq.\eqref{cubic} because the integrands of Eq.\eqref{q1d_wk} and Eq.\eqref{Ep_wk} have the same factor $\bar{\xi}^{+}$ in the denominator. Thus, the integrand of Eq.\eqref{Ep_wk} also has one real pole (bound state $\zeta_0$) and two complex poles (resonance and anti-resonance states, $\zeta_1$ and $\zeta_2$) on the complex frequency plane, which behave as shown in Fig.\ref{fig:exceptional_point}. One can perform the integration in Eq.\eqref{Ep_wk} by analytic continuation of the integration variable into the complex frequency plane. For this case, the complex frequency plane consists of two Riemann sheets as shown in Fig.\ref{fig:residue_paper}. In Fig.\ref{fig:residue_paper}, the bound state exists on the real axis of the first Riemann sheet of the complex frequency plane, whereas the resonance and anti-resonance poles coexist as complex conjugates on the second Riemann sheet. The resonance state appears on the lower half plane, while the anti-resonance state appears on the upper half plane. \\
\indent
For $\tau>0$, we perform a contour deformation as shown in Fig.\ref{fig:residue_paper}. The integral contours $C_-$ and $C_+$ cancel out. Moreover, it is easy to show that the integral contributions $C_{R_{1-}}$ and $C_{R_{2-}}$ become zero in the limit $R\rightarrow \infty$. Then, $E_{p}(0,\tau)$ is composed of only two terms as
\begin{align}
  E_{p}(0,\tau) = E_{p,p}(0,\tau) + E_{p,b}(0,\tau),\label{Ep_two_terms}
\end{align}
where
\begin{align}
 E_{p,p}(0,\tau) &\equiv \lambda E_c \int_{C_{p_-}} d w_k \bar{g}^{+}(w_{\kappa}) e^{-iw_k \tau} + c.c.,\label{Epp}\\
 E_{p,b}(0,\tau) &\equiv \lambda E_c \int_{C_{b_-}} d w_k \bar{g}^{+}(w_{\kappa}) e^{-iw_k \tau} + c.c..\label{Epb}
\end{align}
This means that the pole effect $E_{p,p}(0,\tau)$ and the branch point effect $E_{p,b}(0,\tau)$ are the only contribution to the integration in Eq.\eqref{Ep_wk}.

\subsection{Pole effect}
Since we know the location of the pole $\zeta_1$, we can perform the integral \eqref{Epp} using the residue theorem, thus
\begin{align}
 E_{p,p}(0,\tau) &= -2\pi i \lambda E_c \frac{\sqrt{\zeta_1^2-1} h(\zeta_1) e^{-i\zeta_1 \tau}}{3\zeta_1^2-w_1^2-2} + c.c..\label{Epp_time_evo}
\end{align}
The time evolution of the emitted field at the Van Hove singularity is obtained by substituting Eqs.\eqref{Puiseux_zeta_1^2} and \eqref{Puiseux} into Eq.\eqref{Epp_time_evo} with $w_1=1$, is given by
\begin{align}
 E_{p,p}(0,\tau) &= \frac{4\pi \lambda^{\frac{1}{3}} E_c\left|h(\zeta_1)\right|}{3\sqrt{3}G^{\frac{2}{3}}} \nonumber\\
&\times \sin\left[\left(1+\frac{1}{4}\lambda^{\frac{4}{3}}G^{\frac{4}{3}}\right)\tau + \frac{\pi}{6}+\theta\right]e^{-\frac{\sqrt{3}}{4}\lambda^{\frac{4}{3}}G^{\frac{4}{3}}\tau}.\label{Epp_time_evo_vanhove}
\end{align}
where
\begin{align}
 \left|h(\zeta_1)\right|^2 &= \left\{{\rm Re}[\bar{q}_1(0)]\right\}^2 + \left\{{\rm Im}[\bar{q}_1(0)]\right\}^2\nonumber\\
& + \frac{1}{2}\lambda^{\frac{4}{3}}G^{\frac{4}{3}}\left(\sqrt{3}{\rm Re}[\bar{q}_1(0)]+{\rm Im}[\bar{q}_1(0)]\right){\rm Im}[\bar{q}_1(0)]\nonumber\\
& + \frac{1}{4}\lambda^{\frac{8}{3}}G^{\frac{8}{3}}\left\{{\rm Im}[\bar{q}_1(0)]\right\}^2
\end{align}
The phase $\theta$ satisfies
\begin{align}
 \sin\theta &= \frac{{\rm Im}[\bar{q}_1(0)]-\sqrt{3}{\rm Re}[\bar{q}_1(0)]}{2\left|h(\zeta_1)\right|},\\
 \cos\theta &= \frac{\sqrt{3}{\rm Im}[\bar{q}_1(0)]+{\rm Re}[\bar{q}_1(0)]}{2\left|h(\zeta_1)\right|}.
\end{align}
\subsection{Branch point effect}
Next, we evaluate the branch point effect. The integration path $C_{b_-}$ in Eq.\eqref{Epb} gives the path from the branch point $1$ to $1-iR$ ($R\rightarrow \infty$) in the complex frequency plane (see Fig.\ref{fig:residue_paper}),
\begin{align}
 E_{p,b}(0,\tau) &= \lambda E_c \lim_{R \to \infty} \int_{1}^{1-iR} dw_{\kappa} \bar{g}^{+}(w_{\kappa}) e^{-iw_{\kappa} \tau} + c.c..\label{Epb_wk}
\end{align}
We change the integration variable from $w_{\kappa}$ to $s$ as
\begin{align}
 w_{\kappa} &= 1 - i \frac{s}{\tau},\label{g_to_s}
\end{align}
then, we have
\begin{align}
 E_{p,b}(0,\tau) &=\frac{-i\lambda E_c}{\tau}e^{-i\tau} \int_{0}^{\infty} ds \bar{g}^{+}\left(1-i\frac{s}{\tau}\right)e^{-s} + c.c.,\label{Egs}
\end{align}
where we have taken the limit $R \rightarrow \infty$ and,
\begin{widetext}
\begin{align}
 \bar{g}^{+}\left(1 - i\frac{s}{\tau}\right) &= \frac{w_1\tau^2(\tau - is){\rm{Re}}[\bar{q}_1(0)]+i\tau (\tau - is)^2{\rm{Im}}[\bar{q}_1(0)]}{\sqrt{s}\sqrt{- i 2 \tau - s} \left[(1 - w_1^2)\tau^2 - i2 \tau s -s^2 \right] - i \lambda^2 G^2 w_1\tau^3}.\label{g_s}
\end{align}
\end{widetext}
Since the integrand in Eq.\eqref{Egs} is expressed by known functions, we can evaluate the value of the integral \eqref{Egs} in terms of quadrature by numerical calculations. The results will be shown in the next section.

\begin{figure*}[t!]
\centering
\includegraphics[width=\linewidth]{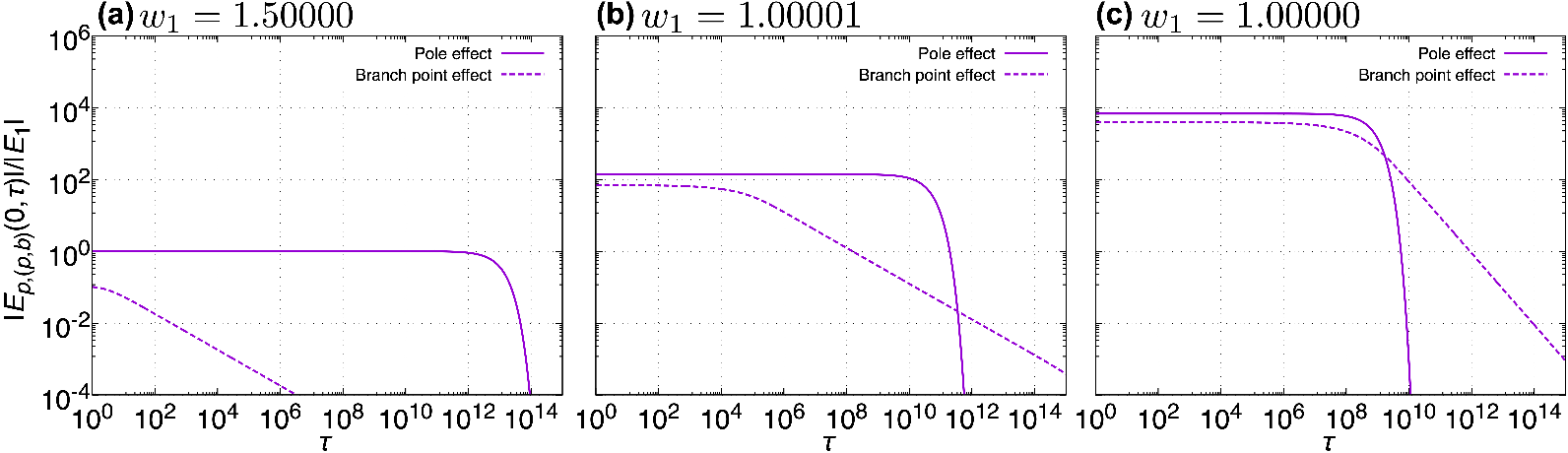}
\caption{Time evolution of the absolute value of the pole and branch point effects: (a), (b) and (c) are represented when $w_1$ is in the Fermi region, between the Fermi region and the Van Hove singularity region, and on the Van Hove singularity, respectively. These absolute values are normalized by $\left| E_1 \right| \equiv \left| E_{p,p}(0,1) \right| $.}
\label{fig:compared_pole_branch}
\end{figure*}
\section{Discussion}
Fig.\ref{fig:compared_pole_branch} shows the numerical results of the absolute value of the pole effect in Eq.\eqref{Epp_time_evo} and the branch point effect in Eq.\eqref{Egs} as functions of the dimensionless cyclotron frequency $w_1$. Fig.\ref{fig:compared_pole_branch}(a), (b) and (c) show the dynamics when $w_1$ is in the Fermi region, between the Fermi region and the Van Hove singularity region (we call this the middle region), and the Van Hove singularity region, respectively. The solid lines represent the pole effect, while the dashed lines represent the branch point effect. In all regions, the pole effect leads to exponential decay, while the branch point effect results in power law decay. \\
\indent
When $w_1$ is in the Fermi region as shown in Fig.\ref{fig:compared_pole_branch}(a), the branch point effect rapidly dies out as the order of the amplitude becomes much smaller than the pole effect, making it difficult to observe experimentally. However, when $w_1$ is in the Van Hove singularity region as shown in Fig.\ref{fig:compared_pole_branch}(c), the branch point effect persists up to the relaxation time of the pole effect with about the same magnitude as the pole effect. This result suggests that the non-Markovian branch point effect is experimentally observable in the Van Hove singularity region. \\
\indent
In each case, the timescale on which most of the decay occurs depends on the resonance state. For the Fermi region $1 \ll w_1$, the function $\bar{\xi}^{\pm}(\zeta)$ in Eq.\eqref{disp_zeta_int} is approximated as
\begin{align}
 \bar{\xi}^{\pm}(\zeta) \simeq \zeta^2 - w_1^2 \pm \frac{\lambda^2G^2w_1}{\sqrt{1-w_1^2}},
\end{align}
where we expand the third term on the right-hand side of Eq.\eqref{disp_zeta_int} in a Taylor series around $w_1$ and keep only the lowest order. Since $\bar{\xi}^{\pm}(\zeta)=0$ gives the pole locations, we can obtain $\zeta^2_1$ in the Fermi region,
\begin{align}
 \zeta^2_1 \simeq w_1^2 -i \frac{\lambda^2G^2w_1}{\sqrt{w_1^2-1}}\label{fermi_squared}.
\end{align}
The resonance eigenvalue $\zeta_1$ in the Fermi region is then given by taking the square root of Eq.\eqref{fermi_squared}, and again expanding to obtain
\begin{align}
 \zeta_1 \simeq w_1 -i \frac{\lambda^2G^2}{2\sqrt{w_1^2-1}}\label{fermi_root},
\end{align}
thus the lifetime of the exponential decay in the Fermi region (Fig.\ref{fig:compared_pole_branch}(a)) is estimated by $1/(\lambda^2G^2 / 2\sqrt{w_1^2-1}) \simeq 9.9 \times 10^{12}$, which is consistent with Fermi's golden rule\footnote{We note that it is well known that the Abraham-Lorentz equation derived by using Li\'enard–Wiechert potentials gives the decay rate of an electron proportional to $\lambda^2$ in addition to the problematic runaway solution. Thus, the Abraham-Lorentz equation can be used only in the Fermi region.}. For the Van Hove singularity region (Fig.\ref{fig:compared_pole_branch}(c)), the lifetime of the exponential decay is estimated from Eq.\eqref{Epp_time_evo_vanhove} as $1/(\sqrt{3}\lambda^{\frac{4}{3}}G^{\frac{4}{3}}/4) \simeq 6.0 \times 10^{8}$. Hence, the timescale is dramatically shortened as a result of the Van Hove singularity.\\
\indent
In our previous work Ref.\cite{PTG05}, our original motivation was partly to find a method to enhance the Markovian exponential decay associated with the resonance pole. However, as revealed here the non-Markovian process associated with the branch-point effect is also (comparably) enhanced in the Van Hove region. Returning to this previous objective, we have found here that by appropriately tuning the cyclotron frequency it is also possible to enhance the Markovian dynamics in absolute terms, while maintaining its relative dominance over the non-Markovian dynamics. This is shown in Fig.\ref{fig:compared_pole_branch}(b) in the middle region.

\section{Concluding Remarks}
In this research, we analyzed the emission from an electron in cyclotron motion in a waveguide using the classical Friedrichs model without relying on perturbation techniques. When the cyclotron frequency is tuned in the vicinity of the cut-off frequency of a waveguide (i.e., in the vicinity of the Van Hove singularity), we found not only that the pole effect is sharply amplified ($10^4$ times) as compared with the case of the Fermi region but also that the branch point effect is amplified to about same order of magnitude. This suggests that the non-Markovian branch point effect is experimentally observable in the Van Hove singularity region. Further, we have found an intermediate case for which the Markovian dynamics are significantly enhanced, but are still dominant compared to the non-Markovian dynamics (see Fig.\ref{fig:compared_pole_branch}(b)). Hence, enhancement of both effects should be experimentally accessible.\\
\indent
Further, we have obtained an essentially self-consistent description of classical radiation damping for the specific case of an electron undergoing cyclotronic motion in a waveguide. In particular, we have implicitly dealt with the self-forces acting on the electron due to its own emitted radiation when we exactly evaluated the integral term in Eq.\eqref{disp_zeta}, which is the equivalent of the self-energy function in the quantum formalism. That is to say, we accounted for the back-reaction on the electron at the equivalent of all levels of perturbation theory through our exact diagonalization procedure. We accomplished this by classicalizing the second quantized Friedrichs model and replacing commutation relations with Poisson brackets for (what turned out to be) an exactly solvable model.\\
\indent
This procedure avoids the well-known problems with obtaining a self-consistent description of radiation damping in classical electrodynamics. In particular, we have not employed time-averaging or other phenomenological approximations that are used to obtain the Abraham-Lorentz equation for the self-force on the electron. We have only employed two significant approximations, neither of which is particularly intrusive: that the interaction with the field is localized on the electron's center of motion and field-field interactions have been neglected. Neither of these approximations results in problems such as pre-acceleration or the runaway solution known to plague the Abraham-Lorentz description. Of course, our model only applies to a specific case; however, one could find other cases where classicalization would be applicable and from that point more detailed physical descriptions could be obtained by the use of perturbation theory.\\
\indent
Finally, in this paper, we have focused on the emission process near the lowest cut-off mode in a rectangular waveguide. In future work, we will study the emission process including the modes with higher-frequency cut-off, which will introduce interesting new effects. Further, we will present results for the emission process from cyclotron motion inside a cylindrical waveguide in which propagating field modes with non-zero angular momentum can be produced. This is the so-called optical vortex. We will show that by tuning the cyclotron frequency to the appropriate waveguide mode, one can select to produce the optical vortex with a range of angular momentum values.

\section*{Acknowledgments}
We thank Dr. Satoshi Tanaka, Dr. Hiroaki Nakamura and Dr. Naomichi Hatano for helpful discussions related to this work. Y.G. acknowledges support from the Japan Society for the Promotion of Science (JSPS) under KAKENHI Grant No. 22K14025 and from the Research Enhancement Strategy Office in the National Institute for Fusion Science. S.G. acknowledges support from JSPS under KAKENHI Grant No. 18K03466.
\appendix

\section{DERIVATION OF THE CLASSICAL FRIEDRICHS MODEL}
In this appendix we derive Eq.\eqref{h_w_dimension} from Eq.\eqref{hehf}. By expanding Eq.\eqref{he}, we obtain
\begin{align}
 H_e &= \frac{1}{2m_e}\left[\bm p + e \bm A_{\rm ex}(\bm r_e)\right]^2 \nonumber\\
&\quad\quad\quad + \frac{e}{m_e}\left[\bm p + e \bm A_{\rm ex}(\bm r_e)\right] \cdot \bm A_{\rm in}(\bm r_e) \nonumber\\
&\quad\quad\quad\quad\quad\quad + \frac{e^2}{2m_e}\bm A_{\rm in}(\bm r_e) \cdot \bm A_{\rm in}(\bm r_e).\label{he_expand}
\end{align}
Here, we will define the unperturbed part $H_e^0$ and perturbed part $H_e^1$ as
\begin{align}
 H_e^0 &\equiv \frac{1}{2m_e}\left[\bm p + e \bm A_{\rm ex}(\bm r_e)\right]^2,\label{he0}\\
 H_e^1 &\equiv \frac{e}{m_e}\left[\bm p + e \bm A_{\rm ex}(\bm r_e)\right] \cdot \bm A_{\rm in}(\bm r_e),\label{he1}
\end{align}
thus,
\begin{align}
 H_e \simeq H_e^0 + H_e^1, \label{he_approx}
\end{align}
where we have neglected the field-field interaction term propotional to the square of $\bm A_{\rm in}(\bm r)$, which is the third term in the right-hand side of Eq.\eqref{he_expand}.\\
\indent
By substituting the explicit form of the symmetric gauge $\bm A_{\rm ex}(\bm r)$ in Eq.\eqref{vp_external} into Eq.\eqref{he0}, $H_e^0$ is given by
\begin{align}
 H_e^0 &= \frac{1}{2m_e}\left\{\left(p_y+\frac{m_e\omega_1}{2}x_e\right)^2 + \left(p_x-\frac{m_e\omega_1}{2}y_e\right)^2\right\},\label{}
\end{align}
where we impose the restriction that the $z$ element of $\bm p$ is not present because it does not affect the emission of field. Since cyclotron motion is circular, it is presumed that $H_e^0$ can be expressed by a harmonic oscillator with one degree of freedom. Under this presumption, we express $H_e^0$ as
\begin{align}
 H_e^0 &=\frac{\omega_1}{2}(P_1^2 + Q_1^2),\label{he0_pq}
\end{align}
where we defined the new variables as
\begin{align}
 P_1 &\equiv \frac{1}{\sqrt{m_e\omega_1}}\left(p_y+\frac{m_e\omega_1}{2}x_e\right),\label{P1}\\
 Q_1 &\equiv \frac{1}{\sqrt{m_e\omega_1}}\left(p_x-\frac{m_e\omega_1}{2}y_e\right).\label{Q1}
\end{align}
For the arbitrary functions $f$ and $g$ with these variables as the arguments, we define the following Poisson bracket:
\begin{align}
 \{f,g\}_1 \equiv \left(\frac{\partial f}{\partial P_1}\frac{\partial g}{\partial Q_1} - \frac{\partial f}{\partial Q_1}\frac{\partial g}{\partial P_1}\right).\label{}
\end{align}
These variables $P_1$ and $Q_1$ satisfy the Poisson brackets below;
\begin{align}
 \{P_1, Q_1\} &= 1,\label{pqpb11}\\
 \{P_1, P_1\} &= 0,\label{pppb11}\\
 \{Q_1, Q_1\} &= 0,\label{qqpb11}
\end{align}
thus, $P_1$ and $Q_1$ are canonical variables. And it is proven that our presumption that $H_e^0$ can be expressed as a harmonic oscillator with one degree of freedom is correct. We note since there were originally four variables ($x_e, y_e, p_x, p_y$), there should also be another pair of canonical variables $P_2$ and $Q_2$, in addition to $P_1$ and $Q_1$. However, $P_2$ and $Q_2$ do not appear in this Hamiltonian, and hence they are constants of the motion in this system.\\
\indent
However, we need to know $P_2$ and $Q_2$ in order to obtain the inverse canonical transformation. Hence we assume $P_2$ and $Q_2$ are a linear combination of $(x_e, y_e, p_x, p_y)$ because $P_1$ and $Q_1$ are also a linear transformation including any of the elements ($x_e, y_e, p_x, p_y$) in Eqs\eqref{P1} and \eqref{Q1}. Namely, $P_2$ and $Q_2$ are assumed as follows:
\begin{align}
 P_2 &= c_1 x_e + c_2 p_x + c_3 y_e + c_4 p_y,\\
 Q_2 &= d_1 x_e + d_2 p_x + d_3 y_e + d_4 p_y.
\end{align}
where $(c_1, c_2, c_3, c_4)$ and $(d_1, d_2, d_3, d_4)$ are unknown coefficients. These coefficients are determined so that $P_2$ and $Q_2$ become canonical variables by using the Poisson brackets;
\begin{align}
 \{P_2, Q_2\}_2 &= 1,\label{pqpbij}\\
 \{P_2, P_2\}_2 &= 0,\label{pppbij}\\
 \{Q_2, Q_2\}_2 &= 0,\label{qqpbij}
\end{align}
where we have defined the following Poisson bracket for the arbitrary functions $f$ and $g$ with these variables as the arguments
\begin{align}
 \{f,g\}_2 \equiv \left(\frac{\partial f}{\partial P_2}\frac{\partial g}{\partial Q_2} - \frac{\partial f}{\partial Q_2}\frac{\partial g}{\partial P_2}\right).\label{}
\end{align}
Thus, we obtain the relations for the coefficients,
\begin{align}
 c_3&=\frac{m_e\omega_1}{2}c_2,\quad c_4=-\frac{2}{m_e\omega_1}c_1,\\
 d_3&=\frac{m_e\omega_1}{2}d_2,\quad d_4=-\frac{2}{m_e\omega_1}d_1,
\end{align}
and
\begin{align}
 c_1d_2 - c_2d_1 = \frac{1}{2}.\label{coeff}
\end{align}
Therefore, $P_2$ and $Q_2$ are given by
\begin{align}
 P_2 &= -\frac{2}{m_e\omega_1}c_1\left(p_y-\frac{m_e\omega_1}{2}x_e\right) + c_2\left(p_x+\frac{m_e\omega_1}{2}y_e\right),\label{p2}\\
 Q_2 &= -\frac{2}{m_e\omega_1}d_1\left(p_y-\frac{m_e\omega_1}{2}x_e\right) + d_2\left(p_x+\frac{m_e\omega_1}{2}y_e\right).\label{q2}
\end{align}
The inverse transformation is given by
\begin{align}
 x_e &= \frac{1}{\sqrt{m_e \omega_1}}P_1 - c_2 Q_2 + d_2 P_2,\label{ctx}\\
 y_e &= -\frac{1}{\sqrt{m_e \omega_1}}Q_1 + \frac{2}{m_e\omega_1}c_1 Q_2 - \frac{2}{m_e\omega_1} d_1 P_2,\label{cty}\\
 p_x &= \frac{\sqrt{m_e\omega_1}}{2} Q_1 + c_1 Q_2 -d_1 P_2,\label{ctpx}\\
 p_y &= \frac{\sqrt{m_e \omega_1}}{2}P_1 + \frac{m_e\omega_1}{2}c_2 Q_2 - \frac{m_e\omega_1}{2} d_2 P_2.\label{ctpy}
\end{align}\\
\indent
By substituting the explicit form of $\bm A_{\rm ex}(\bm r)$ in Eq.\eqref{vp_external} into Eq.\eqref{he1} and carrying out the canonical transformation by using $P_1$ and $Q_1$, $H_e^1$ is given by
\begin{align}
 H_e^1 &= \sqrt{\frac{\omega_1 e^2}{m_e}}\left[Q_1A_{{\rm {in}},x}(\bm r_e) + P_1 A_{{\rm {in}},y}(\bm r_e)\right].\label{he1_ct}
\end{align}
where again only $x$ and $y$ components are present.\\
\indent
Furthermore, we introduce the discrete normal mode by
\begin{align}
 q_1 &\equiv \frac{1}{\sqrt{2}}(P_1 + i Q_1),\label{q1}\\
 q_1^* &\equiv \frac{1}{\sqrt{2}}(P_1 - i Q_1).\label{q1*}
\end{align}
The inverse transformation is given by
\begin{align}
 P_1 &= \frac{1}{\sqrt{2}}(q_1 + q_1^*),\label{p1_inver}\\
 Q_1 &= \frac{-i}{\sqrt{2}}(q_1 - q_1^*).\label{q1_inver}
\end{align}
Thus, the electron part of the Hamiltonian $H_e=H_e^0+H_e^1$ is given by
\begin{align}
 H_e &= \omega_1 q_1^*q_1 \nonumber\\
&- i\sqrt{\frac{\omega_1 e^2}{2m_e}}\left[(q_1-q_1^*)A_{{\rm {in}},x}(\bm r_e) + i(q_1+q_1^*)A_{{\rm {in}},y}(\bm r_e)\right].\label{he_q1}
\end{align}\\
\indent
Also, by substituting the vector potentials \eqref{vp_Ate} and \eqref{vp_Atm} into Eq.\eqref{hf}, the field part of the Hamiltonian $H_f$ is obtained by using the continuous normal mode,
\begin{align}
 H_f=\sum_{o}\sum_{m,n} \int_{k} \omega_{\bm k}{q^{o}_{\bm k}}^*q^{o}_{\bm k}, \label{hf_normal}
\end{align}
where the index of summation $o$ means ${\rm TE}$ or ${\rm TM}$. We note that this Hamiltonian can also be expressed in the form of a harmonic oscillator as follows:
\begin{align}
 H_f = \sum_{o}\sum_{m,n} \int_{k}\left(\frac{1}{2\varepsilon_0}P^{o}_{\bm k}+\frac{1}{2}\varepsilon_0\omega^2_{\bm k}Q^{o}_{\bm k}\right),\label{}
\end{align}
where
\begin{align}
 P^{o}_{\bm k} &\equiv \sqrt{\frac{\varepsilon_0\omega_{\bm k}}{2}}\left(q^{o}_{\bm k} + {q^{o}_{\bm k}}^*\right),\label{}\\
 Q^{o}_{\bm k} &\equiv \frac{-i}{\sqrt{2\varepsilon_0\omega_{\bm k}}}\left(q^{o}_{\bm k} - {q^{o}_{\bm k}}^*\right).\label{}
\end{align}
For the arbitrary functions $f$ and $g$ with these variables as the arguments, we define the following Poisson bracket:
\begin{align}
 \{f,g\}_{\bm k} \equiv \sum_{\bm k} \left(\frac{\partial f}{\partial P^o_{\bm k}}\frac{\partial g}{\partial Q^o_{\bm k}} - \frac{\partial f}{\partial Q^o_{\bm k}}\frac{\partial g}{\partial P^o_{\bm k}}\right).\label{}
\end{align}
Here, $P^{o}_{\bm k}$ and $Q^{o}_{\bm k}$ represent the generalized momentum and the generalized coordinates, respectively, satisfying the Poisson bracket below:
\begin{align}
 \{P^{o}_{\bm k}, Q^{o}_{\bm k'}\}_{\bm k} &= \delta(\bm k - \bm k'),\label{}\\
 \{P^{o}_{\bm k}, P^{o}_{\bm k'}\}_{\bm k} &= 0,\label{}\\
 \{Q^{o}_{\bm k}, Q^{o}_{\bm k'}\}_{\bm k} &= 0.\label{}
\end{align}
The continuous normal modes are given by
\begin{align}
 q^{o}_{\bm k} &= \frac{1}{\sqrt{2\varepsilon_0 \omega_{\bm k}}}\left(P^{o}_{\bm k}+i\varepsilon_0 \omega_{\bm k}Q^{o}_{\bm k}\right),\label{apA_qk}\\
 {q^{o}_{\bm k}}^* &= \frac{1}{\sqrt{2\varepsilon_0 \omega_{\bm k}}}\left(P^{o}_{\bm k}-i\varepsilon_0 \omega_{\bm k}Q^{o}_{\bm k}\right).\label{apA_qk*}
\end{align}
For these Poisson brackets with respect to these canonical variables, we define the following new Poisson brackets:
\begin{align}
 \{f,g\} \equiv \sum_{\alpha'} \left(\frac{\partial f}{\partial P_{\alpha'}}\frac{\partial g}{\partial Q_{\alpha'}} - \frac{\partial f}{\partial Q_{\alpha'}}\frac{\partial g}{\partial P_{\alpha'}}\right),\label{}
\end{align}
where $\alpha'$ is $1$, $2$, or $\bm k$. Thus, introducing the normal mode in this way as Eqs.\eqref{q1}, \eqref{q1*}, \eqref{apA_qk} and \eqref{apA_qk*} yields Eq.\eqref{PB}. Therefore, the full Hamiltonian $H$ is given by using normal modes,
\begin{align}
 H &= \omega_1 q_1^*q_1 + \sum_{o} \sum_{m,n} \int_{k} \omega_{\bm k} {q^{o}_{\bm k}}^{*} q^{o}_{\bm k}\nonumber\\
 &+ \lambda \sum_{o} \sum_{m,n} \int_{k} \left\{(q_1-q_1^*)\left[V^{o}_{\bm k,x}(\bm r_e)q^{o}_{\bm k} - {V^{o}_{\bm k,x}}^*(\bm r_e){q^{o}_{\bm k}}^*\right] \right. \nonumber\\
 & \left. + i(q_1+q_1^*)\left[V^{o}_{\bm k,y}(\bm r_e)q^{o}_{\bm k} - {V^{o}_{\bm k,y}}^*(\bm r_e){q^{o}_{\bm k}}^*\right]\right\},\label{h_normal}
\end{align}
where we have defined the form factors:
\begin{align}
 V_{\bm k,x}^{\rm TE}(\bm r_e) &\equiv i\frac{\sqrt{c^3}}{\sqrt{\pi x_Ly_L}} \sqrt{\frac{\omega_1}{\omega_c}} \frac{n\pi}{y_L} \frac{F^{\rm CS}_{m,n}(x_e,y_e)}{\chi_{m,n}\sqrt{\omega_{\bm k}}}e^{ikz_e},\label{v_x_te}\\
 V_{\bm k, y}^{\rm TE}(\bm r_e) &\equiv -i\frac{\sqrt{c^3}}{\sqrt{\pi x_Ly_L}} \sqrt{\frac{\omega_1}{\omega_c}} \frac{m\pi}{x_L} \frac{F^{\rm SC}_{m,n}(x_e,y_e)}{\chi_{m,n}\sqrt{\omega_{\bm k}}}e^{ikz_e},\label{v_y_te}\\
 V_{\bm k,x}^{\rm TM}(\bm r_e) &\equiv i\frac{\sqrt{c^5}}{\sqrt{\pi x_Ly_L}} \sqrt{\frac{\omega_1}{\omega_c}} \frac{m\pi}{x_L} \frac{k F^{\rm CS}_{m,n}(x_e,y_e)}{\chi_{m,n}\sqrt{\omega_{\bm k}^3}}e^{ikz_e},\label{v_x_tm}\\
 V_{\bm k,y}^{\rm TM}(\bm r_e) &\equiv i\frac{\sqrt{c^5}}{\sqrt{\pi x_Ly_L}} \sqrt{\frac{\omega_1}{\omega_c}} \frac{n\pi}{y_L} \frac{k F^{\rm SC}_{m,n}(x_e,y_e)}{\chi_{m,n}\sqrt{\omega_{\bm k}^3}}e^{ikz_e}.\label{v_y_tm}
\end{align}
In this study, we imposed the restriction that the $z$ element of $\bm p$ is not present because it does not affect the emission of field. Thus the position variable $z_e$ should be replaced by the constant value $z_c$. The quantity $z_c$ denotes the $z$ component of the center of the cyclotron motion $\bm r_c = (x_c,y_c,z_c)$. Therefore, the argument of the above form factors should be represented by $V^{o}_{\bm k,u}(x_e,y_e,z_c)$. Note that the subscript $u$ of $V^{o}_{\bm k,u}(x_e,y_e,z_c)$ stands for $x$ and $y$. \\
\indent
Now, we use the dipole approximation in the form factors, because the radius of cyclotron motion $\bm r_e - \bm r_c$ is much smaller than the width of the waveguide in the both $x$ and $y$ directions:
\begin{align}
 V^{o}_{\bm k,u}(x_e,y_e,z_c) &= V^{o}_{\bm k,u}(\bm r_c) \nonumber\\
 &\!\!\!\!\!\!\!\!\!\!\!\!\!\!\!\!\!\!\!\!\!\!\!\! + \left.\sum_{\chi=1}^{\infty}\frac{1}{\chi!} \left\{\left[(x_e - x_c)\frac{\partial}{\partial x'_e}\right]^{\chi} V^{o}_{\bm k,u}(x'_e,y_e,z_c)\right|_{x'_e = x_c} \right. \nonumber\\
 &\!\!\!\!\!\!\! \left.+ \left.\left[(y_e - y_c)\frac{\partial}{\partial y'_e}\right]^{\chi} V^{o}_{\bm k,u}(x_e,y'_e,z_c)\right|_{y'_e = y_c}\right\}\nonumber\\
 &=V^{o}_{\bm k,u}(\bm r_c) + \mathcal{O}\left(h\left(\frac{x_e-x_c}{x_L},\frac{y_e-y_c}{y_L}\right)\right)\nonumber\\
 &\simeq V^{o}_{\bm k,u}(\bm r_c),\label{v_dipole}
\end{align}
where we have imposed $|(x_e - x_c)/x_L| \ll 1$ and $|(y_e - y_c)/y_L| \ll 1$. Thus, the full Hamiltonian can be approximated as if the interaction between the electron and the field occurs at the rotation center. Therefore we obtain,
\begin{align}
 H &\simeq \omega_1 q_1^*q_1 + \sum_{o} \sum_{m,n} \int_{k} \omega_{\bm k} {q^{o}_{\bm k}}^{*} q^{o}_{\bm k}\nonumber\\
 &+ \lambda \sum_{o} \sum_{m,n} \int_{k} \left\{(q_1-q_1^*)\left[V^{o}_{\bm k,x}(\bm r_c)q^{o}_{\bm k} - {V^{o}_{\bm k,x}}^*(\bm r_c){q^{o}_{\bm k}}^*\right] \right. \nonumber\\
 & \left. + i(q_1+q_1^*)\left[V^{o}_{\bm k,y}(\bm r_c)q^{o}_{\bm k} - {V^{o}_{\bm k,y}}^*(\bm r_c){q^{o}_{\bm k}}^*\right]\right\}.\label{h_before_lowest}
\end{align}
Furthermore, we only consider the ${\rm TE}_{01}$ mode, which is the lowest energy mode. Therefore, the Hamiltonian can be approximated by Eq.\eqref{h_w_dimension}. We note we have replaced some variables in order to avoid heavy notation such as $\omega_{0,1,k} \equiv \omega_k$, $q_{0,1,k}^{\rm TE} \equiv q_{k}$, and $V^{\rm TE}_{0,1,k,x}(\bm r_c) \equiv V_{k}$ in Eq.\eqref{h_w_dimension}.

\section{CLASSICAL BOGOLIUBOV TRANSFORMATION}
In this appendix we derive Eqs.\eqref{h_diago}, \eqref{Q1_bog} and \eqref{Qk_bog}. First, we impose the periodic boundary condition on the Hamiltonian \eqref{h_wo_dimen} using a dimensionless box of size $L$. In this study, the ${\rm TE}_{01}$ mode is taken into consideration, and the Hamiltonian \eqref{h_wo_dimen} is modeled as one-dimensional. Therefore, one-dimensional box normalization is introduced. Then the dimensionless wave-number of the field $\kappa$ is discrete, i.e., $\kappa= 2\pi j/L$ with any integer $j$. Thus Eq.\eqref{h_wo_dimen} reduces to the discretized Hamiltonian $\tilde{H}$ as follows:
\begin{align}
 \tilde{H} &= w_1 \bar{q}_1^* \bar{q}_1 + \sum_{\kappa=-\infty}^{\infty} w_{\kappa} \tilde{q}_{\kappa}^* \tilde{q}_{\kappa} \nonumber\\
& \quad\quad\quad\quad + \lambda \sum_{\kappa=-\infty}^{\infty} \left(\bar{q}_1-\bar{q}_1^*\right)\left(\tilde{V}_{\kappa}\tilde{q}_{\kappa} - \tilde{V}_{\kappa}^*\tilde{q}_{\kappa}^*\right),\label{apB_Hdisc}
\end{align}
where
\begin{align}
 \tilde{q}_{\kappa} &\equiv \sqrt{\frac{2\pi}{L}} \bar{q}_{\kappa},\label{apB_tilde_q}\\
 \tilde{V}_{\kappa} &\equiv \sqrt{\frac{2\pi}{L}} \bar{V}_{\kappa}.\label{apB_tilde_v}
\end{align}
We note $\bar{q}_{\kappa}$ and $\bar{V}_{\kappa}$ have the order of $L^0$. To deal with the continuous wave-number of the field, we will take the limit $L \rightarrow \infty$ in the appropriate stage of calculations. In this limit we have
\begin{align}
 \frac{2\pi}{L}\sum_{\kappa = -\infty}^{\infty} \rightarrow \int_{-\infty}^{\infty} d\kappa ,\quad \frac{L}{2\pi} \delta_{\kappa, \kappa'} \rightarrow \delta(\kappa-\kappa').\label{apB_sumint}
\end{align}\\
\indent
Since Eq.\eqref{apB_Hdisc} is a bilinear Hamiltonian, we can exactly ``diagonalize'' it as
\begin{align}
 \tilde{H} = W_1 \tilde{Q}_1^* \tilde{Q}_1 + \sum_{\kappa} W_{\kappa} \tilde{Q}_{\kappa}^* \tilde{Q}_{\kappa},\label{apB_Hdisc_diag}
\end{align}
where we have abbreviated the summation sign for simplification as
\begin{align}
 \sum_{\kappa} \equiv \sum_{\kappa=-\infty}^{\infty}.\label{apB_sum_abb}
\end{align}
Here $\tilde{Q}_1$ and $\tilde{Q}_{\kappa}$ represent the renormalized normal modes, while $W_1$ and $W_{\kappa}$ denote the renormalized frequencies. The normal modes satisfy the Poisson brackets,
\begin{align}
 \left\{\tilde{Q}_{1}, \tilde{Q}_{1}^*\right\} &= -i,\label{apB_tq1q1}\\
 \left\{\tilde{Q}_{\kappa}, \tilde{Q}_{\kappa'}^*\right\} &= -i\delta_{\kappa,\kappa'},\label{apB_tqkqk}\\
 \left\{\tilde{Q}_{\alpha}, \tilde{Q}_{\beta}\right\} &= 0.\label{apB_tqaqb}
\end{align}
We note that in Eq.\eqref{apB_tqkqk} here, it is the Kronecker delta. The eigenvalue equation for the Liouvillian composed of this diagonalized Hamiltonian \eqref{apB_Hdisc_diag} is the following, with the eigenvalue associated with the renormalized frequency of the Hamiltonian \eqref{apB_Hdisc_diag}:
\begin{align}
 -L_H \tilde{Q}_{\alpha} = W_{\alpha} \tilde{Q}_{\alpha}\label{apB_L_disc}.
\end{align}\\
\indent
For calculating $\tilde{Q}_1$, we assume a linear transformation as follows:
\begin{align}
 \tilde{Q}_1 &= C_{11}\bar{q}_1 + D_{11}\bar{q}_1^* + \sum_{\kappa} \left(C_{1\kappa} \tilde{q}_{\kappa} + D_{1\kappa} \tilde{q}_{\kappa}^* \right),\label{apB_Q1_trans}
\end{align}
where $C_{11}$, $D_{11}$, $C_{1\kappa}$ and $D_{1\kappa}$ are coefficients. By substituting Eq.\eqref{apB_Q1_trans} into Eq.\eqref{apB_L_disc} then comparing the coefficients of $\bar{q}_1$, $\bar{q}^*_1$, $\tilde{q}_{\kappa}$ and $\tilde{q}^*_{\kappa}$ on both sides, we have
\begin{align}
 w_1C_{11} - \lambda\sum_{\kappa}\left(\tilde{V}_{\kappa}^*C_{1\kappa} + \tilde{V}_{\kappa}D_{1\kappa}\right) &= W_1 C_{11},\label{apB_Q1_coeff1}\\
 -w_1D_{11} + \lambda\sum_{\kappa}\left(\tilde{V}_{\kappa}^*C_{1\kappa} + \tilde{V}_{\kappa}D_{1\kappa}\right) &= W_1 D_{11},\label{apB_Q1_coeff2}\\
 w_{\kappa} C_{1\kappa} - \lambda\tilde{V}_{\kappa}\left(C_{11}+D_{11}\right) &= W_1C_{1\kappa},\label{apB_Q1_coeff3}\\
 -w_{\kappa} D_{1\kappa} + \lambda\tilde{V}_{\kappa}\left(C_{11}+D_{11}\right) &= W_1D_{1\kappa}.\label{apB_Q1_coeff4}
\end{align}
From Eqs.\eqref{apB_Q1_coeff1} and \eqref{apB_Q1_coeff2}, we obtain
\begin{align}
 D_{11} = \frac{w_1-W_1}{w_1+W_1}C_{11}.\label{apB_D11C11}
\end{align}
By substituting Eq.\eqref{apB_D11C11} into Eq.\eqref{apB_Q1_coeff3} and Eq.\eqref{apB_Q1_coeff4}, we have
\begin{align}
 C_{1\kappa} &= \frac{2w_1}{w_1+W_1}\frac{\lambda\tilde{V}_{\kappa}}{w_{\kappa}-W_1}C_{11},\label{apB_C1kC11}\\
 D_{1\kappa} &= \frac{2w_1}{w_1+W_1}\frac{\lambda\tilde{V}^*_{\kappa}}{w_{\kappa}+W_1}C_{11}.\label{apB_D1kC11}
\end{align}
Also, by substituting the relationships from Eqs.\eqref{apB_D11C11} to \eqref{apB_D1kC11} into Eq.\eqref{apB_Q1_trans}, we have
\begin{align}
 \tilde{Q}_1 &= \frac{2w_1}{w_1+W_1}C_{11}\left[\frac{w_1+W_1}{2w_1}\bar{q}_1 + \frac{w_1-W_1}{2w_1}\bar{q}^*_1 \right.\nonumber\\
 &\quad\quad\quad \left.+ \sum_{\kappa}\left(\frac{\lambda\tilde{V}_{\kappa}}{w_{\kappa}-W_1}\tilde{q}_{\kappa} + \frac{\lambda\tilde{V}^*_{\kappa}}{w_{\kappa}+W_1}\tilde{q}^*_{\kappa} \right)\right].\label{apB_Q1C11}
\end{align}
Furthermore, by substituting Eqs.\eqref{apB_C1kC11} and \eqref{apB_D1kC11} into Eq.\eqref{apB_Q1_coeff1}, we obtain a transcendental equation that provides a specific representation of $W_1$:
\begin{align}
 \tilde{\xi}_{1}(W_1) = 0,\label{apB_W1_transQ1}
\end{align}
where
\begin{align}
 \tilde{\xi}_{1}(W) \equiv W^2 - w_1^2 - 4\lambda^2 w_1\sum_{\kappa}\frac{w_{\kappa}|\tilde{V}_{\kappa}|^2}{W^2-w_{\kappa}^2}.\label{apB_transQ1_form}
\end{align}
The coefficient $C_{11}$ can be identified from Poisson brackets. By substituting $\tilde{Q}_1$ and its complex conjugate into Eq.\eqref{apB_tq1q1}, we obtain
\begin{align}
 C_{11} = \frac{w_1+W_1}{2w_1}\tilde{N}_1,\label{apB_C11}
\end{align}
where
\begin{align}
 \tilde{N}_1 \equiv \left[\left.\frac{1}{2w_1}\frac{d\tilde{\xi}_{1}(\zeta)}{d\zeta}\right|_{\zeta=W_1}\right]^{-\frac{1}{2}}.\label{apB_tilde_N1}
\end{align}
Here we take the limit of $L \rightarrow \infty$. In Eq.\eqref{apB_transQ1_form}, $|\tilde{V}_{\kappa}|^2$ in Eq.\eqref{apB_tilde_v} has the order of $L^{-1}$, so the summation sign is replaced by an integral symbol, then Eq.\eqref{apB_transQ1_form} reduces to Eq.\eqref{disp_zeta}. This integral can be exactly performed, and from Eq.\eqref{apB_W1_transQ1}, we obtain $W_1=\tilde{w}_1$, which is a real eigenvalue of the Liouvillian on the first Riemann sheet. This means that there is a frequency shift for $W_1$ from $w_1$. Additionally, $\tilde{N}_1$ also reduces to $\bar{N}_1$ in Eq.\eqref{N1}, thus Eq.\eqref{apB_Q1C11} reduces to Eq.\eqref{Q1_bog} on the limit of $L \rightarrow \infty$.
\\
\indent
For $\tilde{Q}_{\kappa}$, we can derive it using the same algebraic calculations as when we derived $\tilde{Q}_1$. In that case, $\tilde{Q}_{\kappa}$ is obtained by replacing $W_1$ with $W_{\kappa}$ in Eq.\eqref{apB_Q1C11}. However, for the coefficient of $\tilde{q}_{\kappa}$, if $W_{\kappa} = w_{\kappa}$, this point becomes a divergence point under the $\kappa$ integral. Therefore, this point is considered separately in advance. In fact, through the following discussion, we will find that the frequency shift of $W_{\kappa}$ is given by $W_{\kappa} = w_{\kappa} + \mathcal{O}(L^{-1})$, and it becomes evident that there is no frequency shift in the field as the limit of $L \rightarrow \infty$. Thus we assume such a linear transformations as follows:
\begin{align}
 \tilde{Q}_{\kappa} &= C_{\kappa 1}\bar{q}_1 + D_{\kappa 1}\bar{q}_1^* + C_{\kappa \kappa} \tilde{q}_{\kappa} \nonumber\\
 &\quad\quad\quad\quad\quad\quad + {\sum_{\kappa'}}' C'_{\kappa \kappa'} \tilde{q}_{\kappa'} + \sum_{\kappa''} D_{\kappa \kappa''} \tilde{q}_{\kappa''}^* ,\label{apB_Qk_trans}
\end{align}
where $C_{\kappa 1}$, $D_{\kappa 1}$, $C_{\kappa \kappa}$, $C'_{\kappa \kappa'}$ and $D_{\kappa \kappa''}$ are coefficients and the summation symbol with prime means that $\kappa'=\kappa$ is eliminated as
\begin{align}
 {\sum_{\kappa'}}' \equiv \sum_{\substack{\kappa' \\ \kappa'\neq \kappa}}.\label{apB_sum_kdash}
\end{align}
By substituting Eq.\eqref{apB_Qk_trans} into Eq.\eqref{apB_L_disc} then comparing the coefficients of $\bar{q}_1$, $\bar{q}^*_1$, $\tilde{q}_{\kappa}$, $\tilde{q}_{\kappa'}$ and $\tilde{q}^*_{\kappa}$ on both sides, we have
\begin{align}
 w_1C_{\kappa 1} - \lambda\left(\tilde{V}_{\kappa}^*C_{\kappa \kappa} + {\sum_{\kappa'}}'\tilde{V}_{\kappa'}^* C'_{\kappa \kappa'} + \sum_{\kappa''}\tilde{V}_{\kappa''}D_{\kappa \kappa''}\right)& \nonumber\\
 & \!\!\!\!\!\!\!\!\!\!\!\!\!\!\!\!\!\!\!\!\!\!\!\!\!\!\!\!\!\!\!\!\!\!\!\!\!\!\!\!\!\!\!\!\!\!\!\!\!\!\!\!= W_{\kappa}C_{\kappa 1},\label{apB_Qk_coeff1}\\
 -w_1D_{\kappa 1} + \lambda\left(\tilde{V}_{\kappa}^*C_{\kappa \kappa} + {\sum_{\kappa'}}'\tilde{V}_{\kappa'}^* C'_{\kappa \kappa'} + \sum_{\kappa''}\tilde{V}_{\kappa''}D_{\kappa \kappa''}\right)& \nonumber\\
 & \!\!\!\!\!\!\!\!\!\!\!\!\!\!\!\!\!\!\!\!\!\!\!\!\!\!\!\!\!\!\!\!\!\!\!\!\!\!\!\!\!\!\!\!\!\!\!\!\!\!\!\!= W_{\kappa}D_{\kappa 1},\label{apB_Qk_coeff2}
\end{align}
\vspace{-9mm}
\begin{align}
 w_{\kappa}C_{\kappa \kappa} - \lambda\tilde{V}_{\kappa}\left(C_{\kappa 1} + D_{\kappa 1}\right) &= W_{\kappa}C_{\kappa \kappa},\label{apB_Qk_coeff3}\\
 -w_{\kappa''}D_{\kappa \kappa''} + \lambda\tilde{V}^*_{\kappa''}\left(C_{\kappa 1} + D_{\kappa 1}\right) &= W_{\kappa}D_{\kappa \kappa''},\label{apB_Qk_coeff4}\\
 w_{\kappa'}C'_{\kappa \kappa'} - \lambda\tilde{V}_{\kappa'}\left(C_{\kappa 1} + D_{\kappa 1}\right) &= W_{\kappa}C'_{\kappa \kappa'}.\label{apB_Qk_coeff5}
\end{align}
From Eqs.\eqref{apB_Qk_coeff1} and \eqref{apB_Qk_coeff2}, we obtain
\begin{align}
 D_{\kappa 1} = \frac{w_1-W_{\kappa}}{w_1+W_{\kappa}}C_{\kappa 1}.\label{apB_Dk1Ck1}
\end{align}
By substituting $D_{\kappa \kappa''}$ of Eq.\eqref{apB_Qk_coeff4} with Eq.\eqref{apB_Dk1Ck1} into Eq.\eqref{apB_Qk_coeff1} and $C'_{\kappa \kappa'}$ of Eq.\eqref{apB_Qk_coeff5} with Eq.\eqref{apB_Dk1Ck1} into Eq.\eqref{apB_Qk_coeff2}, we have
\begin{align}
 C_{\kappa 1} &= -\frac{w_1+W_{\kappa}}{\tilde{\xi}_{2}(W_{\kappa})}\lambda\tilde{V}_{\kappa}^*C_{\kappa \kappa},\label{apB_Ck1Ckk}\\
 D_{\kappa 1} &= -\frac{w_1-W_{\kappa}}{\tilde{\xi}_{2}(W_{\kappa})}\lambda\tilde{V}_{\kappa}^*C_{\kappa \kappa},\label{apB_Dk1Ckk}
\end{align}
where
\begin{align}
 \tilde{\xi}_{2}(W_{\kappa}) \equiv W_{\kappa}^2 - w_1^2 + \frac{2w_1\lambda^2|\tilde{V}_{\kappa}|^2}{W_{\kappa}^2-w_{\kappa'}^2} - 4\lambda^2 w_1 {\sum_{\kappa'}}' \frac{w_{\kappa'}|\tilde{V}_{\kappa'}|^2}{W_{\kappa}^2-w_{\kappa'}^2}.\label{apB_xi_Qk}
\end{align}
By substituting Eq.\eqref{apB_Ck1Ckk} with Eq.\eqref{apB_Dk1Ck1} into Eq.\eqref{apB_Qk_coeff4} and Eq.\eqref{apB_Dk1Ckk} with Eq.\eqref{apB_Dk1Ck1} into Eq.\eqref{apB_Qk_coeff5}, we have
\begin{align}
 C'_{\kappa \kappa'} &= \frac{-2w_1\lambda\tilde{V}_{\kappa}^*}{\tilde{\xi}_{2}(W_{\kappa})}\frac{\lambda\tilde{V}_{\kappa'}}{w_{\kappa'}-W_{\kappa}}C_{\kappa \kappa},\label{apB_CdashkkCCkk}\\
 D_{\kappa \kappa''} &= \frac{-2w_1\lambda\tilde{V}_{\kappa}^*}{\tilde{\xi}_{2}(W_{\kappa})}\frac{\lambda\tilde{V}_{\kappa''}^*}{w_{\kappa''}+W_{\kappa}}C_{\kappa \kappa}.\label{apB_DkkdashdashCkk}
\end{align}
Also, by substituting these relationships Eqs.\eqref{apB_Ck1Ckk}, \eqref{apB_Dk1Ckk}, \eqref{apB_CdashkkCCkk} and \eqref{apB_DkkdashdashCkk} into Eq.\eqref{apB_Qk_trans}, we have
\begin{align}
 \tilde{Q}_{\kappa} &= C_{\kappa \kappa}\left[\tilde{q}_{\kappa} - \frac{2w_1\lambda\tilde{V}_{\kappa}^*}{\tilde{\xi}_{2}(W_{\kappa})}\left(\frac{W_{\kappa}+w_1}{2w_1}\bar{q}_1 - \frac{W_{\kappa}-w_1}{2w_1}\bar{q}_1^* \right.\right.\nonumber\\
&\quad\quad\quad\left.\left.+ {\sum_{\kappa'}}'\frac{\lambda\tilde{V}_{\kappa'}}{w_{\kappa'}-W_{\kappa}}\tilde{q}_{\kappa'} + \sum_{\kappa''}\frac{\lambda\tilde{V}^*_{\kappa''}}{w_{\kappa''}+W_{\kappa}}\tilde{q}_{\kappa''}\right)\right].\label{apB_Qk_Ckk}
\end{align}
The coefficient $C_{\kappa \kappa}$ can be identified from Poisson brackets. Substituting $\tilde{Q}_{\kappa}$ and its complex conjugate into Eq.\eqref{apB_tqkqk}, we obtain
\begin{align}
 C_{\kappa \kappa}C_{\kappa \kappa}^* + \mathcal{O}(L^{-1}) = 1.\label{apB_Ckk_order}
\end{align}
Here again we take the limit of $L \rightarrow \infty$. Then, the second term on the left-hand side of Eq.\eqref{apB_Ckk_order} vanishes due to the order of $L^{-1}$ in this limit. Thus we put
\begin{align}
 C_{\kappa \kappa} = C_{\kappa \kappa}^* = 1.\label{apB_Ckk_1}
\end{align}
Furthermore, by substituting Eqs.\eqref{apB_Ck1Ckk} and \eqref{apB_Dk1Ckk} into Eq.\eqref{apB_Qk_coeff3}, we obtain
\begin{align}
 W_{\kappa} = w_{\kappa} + \frac{2\omega_1\lambda^2|\tilde{V}_{\kappa}|^2}{\tilde{\xi}_{2}(W_{\kappa})}.\label{apB_Wk_noshift}
\end{align}
In Eq.\eqref{apB_Wk_noshift}, under the limit of $L \rightarrow \infty$, it becomes $W_{\kappa}=w_{\kappa}$ because $|\tilde{V}_{\kappa}|^2$ in Eq.\eqref{apB_tilde_v} has the order of $L^{-1}$. This means that the renormalized field frequencies are understood to remain unchanged from the original field frequencies. Finally, in Eq.\eqref{apB_xi_Qk}, $|\tilde{V}_{\kappa}|^2$ in Eq.\eqref{apB_tilde_v} has the order of $L^{-1}$, so the summation sign is replaced by an integral symbol and the third term of the right-hand side vanished, then Eq.\eqref{apB_xi_Qk} reduces to Eq.\eqref{disp_zeta}. Therefore, Eq.\eqref{apB_Qk_Ckk} is reduced to Eq.\eqref{Qk_bog} on the limit of $L \rightarrow \infty$. Here, the direction of analytic continuation is chosen such that the description of motion is more mathematically concise for $\tau>0$ \cite{PPT91}.\\
\indent
By discretizing in this manner, one can understand that in the Hamiltonian after diagonalization under the limit $L \rightarrow \infty$, the renormalized field frequencies become the same as those of the original field frequencies. On the other hand, the frequency of renormalized particle changes after diagonalization. At the end, as a consequence of this limit of $L \rightarrow \infty$, Eq.\eqref{apB_Hdisc_diag} reduces to Eq.\eqref{h_diago}.


\begin{thebibliography}{99}
\bibitem{Jackson98} J. D. Jackson,
Classical Electrodynamics, 3rd Ed.,
(Wiley, New Jersey, 1998).

\bibitem{POP03} T. Petrosky, G. Ordonez, and I. Prigogine,
Radiation damping in classical systems: The role of nonintegrability,
\href{https://doi.org/10.1103/PhysRevA.68.022107}
{Phys. Rev. A, \textbf{68}, 022107 (2003)}.

\bibitem{Friedrichs48} K. Friedrichs,
On the perturbation of continuous spectra,
\href{https://doi.org/10.1002/cpa.3160010404}
{Commun. Pure Appl. Math., \textbf{1}, 361 (1948)}.

\bibitem{KPPP00} E. Karpov, I. Prigogine, T. Petrosky, and G. Pronko,
Friedrichs model with virtual transitions. Exact solution and indirect spectroscopy,
\href{https://doi.org/10.1063/1.533125}
{J. Math. Phys., \textbf{41}, (2000) 118-131}.

\bibitem{PPT91} T. Petrosky, I. Prigogine, and S. Tasaki,
Quantum theory of non-integrable systems,
\href{https://doi.org/10.1016/0378-4371(91)90257-D}
{Phys. A: Stat., \textbf{173}, (1991)}.

\bibitem{GP11} M. Gadella and G. P. Pronko,
The Friedrichs model and its use in resonance phenomena,
\href{https://doi.org/10.1002/prop.201100038}
{Fortschr. Phys., \textbf{59}, 795-859 (2011)}.

\bibitem{Mahan90} G. D. Mahan,
Many-Particle Physics, 2nd Ed.,
(Plenum, New York, 1990).

\bibitem{Van-Hove53} L. Van Hove,
The Occurrence of Singularities in the Elastic Frequency Distribution of a Crystal,
\href{https://doi.org/10.1103/PhysRev.89.1189}
{Phys. Rev., \textbf{89}, 1189 (1953)}.

\bibitem{Kleppner81} D. Kleppner,
Inhibited Spontaneous Emission,
\href{https://doi.org/10.1103/PhysRevLett.47.233}
{Phys. Rev. Lett. \textbf{47}, 233 (1981)}.

\bibitem{KKS94} A. G. Kofman, G. Kurizki, and B. Sherman,
Spontaneous and induced atomic decay in photonic band structures,
\href{https://www.tandfonline.com/doi/abs/10.1080/09500349414550381}
{J. Mod. Opt., \textbf{41}, 353 (1994)}.

\bibitem{JQ94} S. John and T. Quang,
Spontaneous emission near the edge of a photonic band gap,
\href{https://journals.aps.org/pra/abstract/10.1103/PhysRevA.50.1764}
{Phys. Rev. A, \textbf{50}, 1764 (1994)}.

\bibitem{LNNB00} P. Lambropoulos, G. M. Nikolopoulos, T. R. Nielson, and S. Bay,
Fundamental quantum optics in structured reservoirs,
\href{https://iopscience.iop.org/article/10.1088/0034-4885/63/4/201/pdf}
{Rep. Prog. Phys., \textbf{63}, 455 (2000)}.

\bibitem{PTG05} T. Petrosky, C. O. Ting, and S. Garmon,
Strongly Coupled Matter Field and Nonanalytic Decay Rate of Dipole Molecules in a Waveguide,
\href{https://doi.org/10.1103/PhysRevLett.94.043601}
{Phys. Rev. Lett., \textbf{94}, 043601 (2005)}.

\bibitem{Khalfin58} L. A. Khalfin,
Contribution to the decay theory of a quasi-stationary state,
Sov. Phys. JETP, \textbf{6}, 1053 (1958).

\bibitem{FGR78} L. Fonda, G. C. Ghirardi, and A. Rimini,
Decay theory of unstable quantum systems,
\href{https://iopscience.iop.org/article/10.1088/0034-4885/41/4/003}
{Rep. Prog. Phys., \textbf{41} 587 (1978)}.

\bibitem{RHM06} C. Rothe, S. I. Hintschich, and A. P. Monkman,
Violation of the Exponential-Decay Law at Long Times,
\href{https://doi.org/10.1103/PhysRevLett.96.163601}
{Phys. Rev. Lett., \textbf{96}, 163601 (2006)}.

\bibitem{TMMS10}
E. Torrontegui, J. G. Muga, J. Martorell, and D. W. L. Spring,
Quantum Decay at Long Times,
\href{https://www.sciencedirect.com/science/article/pii/S0065327610600093}
{Adv. Quant. Chem., \textbf{60}, 485 (2010)}.

\bibitem{GPSS13} S. Garmon, T. Petrosky, L. Simine and D. Segal,
Amplification of non-Markovian decay due to bound state absorption into continuum,
\href{https://doi.org/10.1002/prop.201200077}
{Fortschr. Phys., \textbf{61}, 261 (2013)}.

\bibitem{CPFSMNPO19} A. Crespi, F. V. Pepe, P. Facchi, F. Sciarrino, P. Mataloni, H. Nakazato, S. Pascazio, and R. Osellame,
Experimental Investigation of Quantum Decay at Short, Intermediate, and Long Times via Integrated Photonics,
\href{https://doi.org/10.1103/PhysRevLett.122.130401}
{Phys. Rev. Lett., \textbf{122}, 130401 (2019)}.

\bibitem{JMSST05} T. Jittoh, S. Matsumoto, J. Sato, Y. Sato, and K. Takeda,
Nonexponential decay of an unstable quantum system: Small-Q-value s-wave decay,
\href{https://journals.aps.org/pra/abstract/10.1103/PhysRevA.71.012109}
{Phys. Rev. A, \textbf{71}, 012109 (2005)}.

\bibitem{GCV06} G. Garc\'ia-Calder\'on and J. Villavicencio,
Full-time nonexponential decay in double-barrier quantum structures,
\href{https://journals.aps.org/pra/abstract/10.1103/PhysRevA.73.062115}
{Phys. Rev. A, \textbf{73}, 062115 (2006)}.

\bibitem{DBP08}
A. D. Dente, R. A. Bustos-Mar\`{u}n, and H. M. Pastawski,
Dynamical regimes of a quantum SWAP gate beyond the Fermi golden rule,
\href{https://doi.org/10.1103/PhysRevA.78.062116}
{Phys. Rev. A, \textbf{78}, 062116 (2008)}.

\bibitem{GNOS19} S. Garmon, K. Noba, G. Ordonez, and D. Segal,
Non-Markovian dynamics revealed at a bound state in the continuum,
\href{https://doi.org/10.1103/PhysRevA.99.010102}
{Phys. Rev. A, \textbf{99}, 010102(R) (2019)}.

\bibitem{GOH21} S. Garmon, G. Ordonez, and N. Hatano,
Anomalous-order exceptional point and non-Markovian Purcell effect at threshold in one-dimensional continuum systems,
\href{https://doi.org/10.1103/PhysRevResearch.3.033029}
{Phys. Rev. Res., \textbf{3}, 033029 (2021)}.


\end{thebibliography}
\end{document}